\documentclass[usenatbib,iop,numberedappendix, letterpaper]{aeb_emulateapj_2010}
\usepackage{amsmath}
\usepackage{amssymb}
\usepackage{xspace}
\usepackage[normalem]{ulem}
\usepackage{mathrsfs}
\usepackage{natbib}

\usepackage[usenames,dvips]{color}

\def\del#1{{}}

\newcommand\bcdot{\ensuremath{%
  \mathchoice%
   {\mskip\thinmuskip\lower0.2ex\hbox{\scalebox{1.5}{$\cdot$}}\mskip\thinmuskip}}%
   {\mskip\thinmuskip\lower0.2ex\hbox{\scalebox{1.5}{$\cdot$}}\mskip\thinmuskip}%
   {\lower0.3ex\hbox{\scalebox{1.2}{$\cdot$}}}%
   {\lower0.3ex\hbox{\scalebox{1.2}{$\cdot$}}}%
}

\def\s{{\rm s}} 

\def\m{{\rm m}} 
\def\cm{{\rm c}\m} 

\def\K{{\rm K}} 

\SetSymbolFont{symbols}{bold}{OMS}{cmsy}{b}{n}
\DeclareSymbolFont{bmisymbols}{OML}{cmm}{b}{it}
\DeclareMathSymbol{\balpha}{0}{bmisymbols}{"0B}
\DeclareMathSymbol{\bbeta}{0}{bmisymbols}{"0C}
\DeclareMathSymbol{\bgamma}{0}{bmisymbols}{"0D}
\DeclareMathSymbol{\bdelta}{0}{bmisymbols}{"0E}
\DeclareMathSymbol{\bepsilon}{0}{bmisymbols}{"0F}
\DeclareMathSymbol{\bzeta}{0}{bmisymbols}{"10}
\DeclareMathSymbol{\boldeta}{0}{bmisymbols}{"11}
\DeclareMathSymbol{\btheta}{0}{bmisymbols}{"12}
\DeclareMathSymbol{\biota}{0}{bmisymbols}{"13}
\DeclareMathSymbol{\bkappa}{0}{bmisymbols}{"14}
\DeclareMathSymbol{\blambda}{0}{bmisymbols}{"15}
\DeclareMathSymbol{\bmu}{0}{bmisymbols}{"16}
\DeclareMathSymbol{\bnu}{0}{bmisymbols}{"17}
\DeclareMathSymbol{\bxi}{0}{bmisymbols}{"18}
\DeclareMathSymbol{\bpi}{0}{bmisymbols}{"19}
\DeclareMathSymbol{\brho}{0}{bmisymbols}{"1A}
\DeclareMathSymbol{\bsigma}{0}{bmisymbols}{"1B}
\DeclareMathSymbol{\btau}{0}{bmisymbols}{"1C}
\DeclareMathSymbol{\bupsilon}{0}{bmisymbols}{"1D}
\DeclareMathSymbol{\bphi}{0}{bmisymbols}{"1E}
\DeclareMathSymbol{\bchi}{0}{bmisymbols}{"1F}
\DeclareMathSymbol{\bpsi}{0}{bmisymbols}{"20}
\DeclareMathSymbol{\bomega}{0}{bmisymbols}{"21}
\DeclareMathSymbol{\bvarepsilon}{0}{bmisymbols}{"22}
\DeclareMathSymbol{\bvartheta}{0}{bmisymbols}{"23}
\DeclareMathSymbol{\bvarpi}{0}{bmisymbols}{"24}
\DeclareMathSymbol{\bvarrho}{0}{bmisymbols}{"25}
\DeclareMathSymbol{\bvarsigma}{0}{bmisymbols}{"26}
\DeclareMathSymbol{\bvarphi}{0}{bmisymbols}{"27}

\def\Fermi{{\em Fermi}\xspace}
\newcommand {\apgt} {\ {\raise-.5ex\hbox{$\buildrel>\over\sim$}}\ }
\newcommand {\aplt} {\ {\raise-.5ex\hbox{$\buildrel<\over\sim$}}\ }

\newcommand{\mathbfit}[1]{\textbf{\textit{#1}}}

\newcommand{\eps}{\varepsilon}
\newcommand{\rmn}{\mathrm}
\newcommand{\CR}{\rmn{CR}}
\newcommand{\B}{\mathcal{B}}
\newcommand{\M}{\mathcal{M}}
\newcommand{\calL}{\mathcal{L}}
\newcommand{\vel}{\upsilon}
\newcommand{\bvel}{\bupsilon}
\newcommand{\bra}{\langle}
\newcommand{\ket}{\rangle}

\journalinfo{Draft version \today}

\begin{document}

\title{Toward a comprehensive model for feedback by active galactic nuclei:\\
  new insights from M87 observations by LOFAR, {\em FERMI}, and H.E.S.S.}

\author{
Christoph Pfrommer
}
\altaffiltext{}{Heidelberg Institute for Theoretical Studies, Schloss-Wolfsbrunnenweg 35, D-69118 Heidelberg, Germany}

\shorttitle{A Comprehensive Model for AGN feedback}
\shortauthors{Pfrommer, C.}

\begin{abstract}
  \noindent
  Feedback by active galactic nuclei (AGNs) appears to be critical in balancing
  radiative cooling of the low-entropy gas at the centers of galaxy clusters and
  in mitigating the star formation of elliptical galaxies. New observations of
  M87 enable us to put forward a comprehensive model for the physical heating
  mechanism. Low-frequency radio observations by LOFAR revealed the absence of
  {\em fossil} cosmic ray (CR) electrons in the radio halo surrounding M87. This
  puzzle can be resolved by accounting for the CR release from the radio lobes
  and the subsequent mixing of CRs with the dense ambient intracluster gas,
  which thermalizes the electrons on a timescale similar to the radio halo age
  of 40 Myr. Hadronic interactions of similarly injected CR protons with the
  ambient gas should produce an observable gamma-ray signal in accordance with
  the steady emission of the low state of M87 detected by \Fermi and
  H.E.S.S. Hence, we normalize the CR population to the gamma-ray emission,
  which shows the same spectral slope as the CR injection spectrum probed by
  LOFAR, thereby supporting a common origin. We show that CRs, which stream at
  the Alfv{\'e}n velocity with respect to the plasma rest frame, heat the
  surrounding thermal plasma at a rate that balances that of radiative cooling
  {\em on average} at each radius. However, the resulting global thermal
  equilibrium is locally unstable and allows for the formation of the observed
  cooling multi-phase medium through thermal instability. Provided that CR heating
  balances cooling during the emerging ``cooling flow,'' the collapse of the
  majority of the gas is halted around 1 keV---in accordance with X-ray data. We
  show that both the existence of a temperature floor and the similar radial
  scaling of the heating and cooling rates are generic predictions of the CR
  heating model.
\end{abstract}

\keywords{
cosmic rays ---
galaxies: active ---
galaxies: clusters: intracluster medium ---
galaxies: individual (M87) ---
gamma rays: galaxies: clusters ---
radio continuum: galaxies}

\maketitle

\section{Introduction} \label{sec:I}

The central cooling time of the X-ray emitting gas in galaxy clusters and groups
is bimodally distributed; approximately half of all systems have radiative
cooling times of less than 1~Gyr and establish a population of cool core (CC)
clusters \citep{2009ApJS..182...12C,2010A&A...513A..37H}. In the absence of any
heating process, these hot gaseous atmospheres are expected to cool and to form
stars at rates up to several hundred $M_\odot~\rmn{yr}^{-1}$ \citep[see][for a
review]{2006PhR...427....1P}. Instead, the observed gas cooling and star
formation rates are reduced to levels substantially below those expected from
unimpeded cooling flows. High-resolution {\em Chandra} and {\em XMM-Newton}
observations show that while the gas temperature is smoothly declining toward
the center, there is a lack of emission lines from gas at a temperature below
about 1 keV \citep[see the multi-temperature models of][]{2010A&A...513A..37H}.
Apparently, radiative cooling is offset by a yet to be identified heating
process that is associated with the active galactic nucleus (AGN) jet-inflated
radio lobes that are co-localized with the cavities seen in the X-ray maps. The
interplay of cooling gas, subsequent star formation, and nuclear activity appears
to be tightly coupled to a self-regulated feedback loop \citep[for reviews,
see][]{2007ARA&A..45..117M, 2012NJPh...14e5023M}. Feedback is considered to be a
crucial if not unavoidable process of the evolution of galaxies and supermassive
black holes \citep{2006MNRAS.365...11C, 2013MNRAS.428.2966P}.  While the
energetics of AGN feedback is sufficient to balance radiative cooling, it is
unable to transform CC into non-CC clusters on the buoyancy timescale due to the
weak coupling between the mechanical energy to the cluster gas
\citep{2012ApJ...752...24P}. Matching this observational fact has proven to be
difficult in cosmological simulations of AGN feedback that include realistic
metal-line cooling \citep{2011MNRAS.417.1853D}.

Several physical processes have been proposed to be responsible for the heating,
including an inward transport of heat from the hot outer cluster regions
\citep{2001ApJ...562L.129N}, turbulent mixing \citep{2003ApJ...596L.139K},
redistribution of heat by buoyancy-induced turbulent convection
\citep{2007ApJ...671.1413C, 2009ApJ...699..348S}, and dissipation of mechanical
heating by outflows, lobes, or sound waves from the central AGN
\citep[e.g.,][]{2001ApJ...554..261C, 2002Natur.418..301B, 2002ApJ...581..223R,
  2004ApJ...611..158R,2012MNRAS.424..190G}. The total energy required to inflate
a cavity of volume $V$ in an atmosphere of thermal pressure $P_\rmn{th}$ is
equal to its enthalpy
$E_\rmn{cav}=\gamma/(\gamma-1)\,P_\rmn{th}V=4\,P_\rmn{th}V$, assuming a
relativistic equation of state with $\gamma=4/3$ within the lobes.
Initially, the partitioning to mechanical energy resulting from the volume
work done on the surroundings only makes use of one-quarter of the total
available enthalpy. Only a fraction thereof is dissipated in the central
regions, which have the smallest cooling time (since most of the temperature
increase caused by weak shocks is provided adiabatically).

There are two possibilities how to dissipate the remaining part of
$3\,P_\rmn{th}V$ that is stored in the internal energy of the (presumably)
relativistic particle population. (1) If these lobes rise buoyantly and
unimpeded over several pressure scale heights, the relativistic lobe filling
does $PdV$ work on the surroundings, which expands the lobe and transfers the
internal lobe energy adiabatically to mechanical energy of the ambient
intracluster medium (ICM). (2) However, {\em if} those cosmic rays (CRs) were
mixed into the ambient gas early on during their buoyant rise, there would be a
promising process to transfer CR into thermal energy.  Fast-streaming CRs along
the magnetic field excite Alfv\'en waves through the ``streaming instability''
\citep{1969ApJ...156..445K}. Scattering off this wave field effectively limits
the bulk speed of CRs. Damping of these waves transfers CR energy and momentum
to the thermal gas at a rate that scales with the CR pressure gradient. Hence,
this could provide an efficient means of suppressing the cooling catastrophe of
cooling cores \citep[CCs;][]{1991ApJ...377..392L, 2008MNRAS.384..251G,
  2011A&A...527A..99E, 2012ApJ...746...53F, 2013MNRAS.434.2209W}, but only {\em
  if} the CR pressure gradient is sufficiently
large.\footnote{\citet{1991ApJ...377..392L} only account for slow diffusive CR
  transport that was calibrated to interstellar medium conditions. As a result,
  their models become CR pressure dominated at the cluster center, which yields
  too flat thermal pressure profiles in comparison to observations.  Considering
  instead CR transport by the buoyant lobes observed in {\em Chandra} X-ray
  images, which are subsequently disrupted by Rayleigh--Taylor or
  Kelvin--Helmholtz instabilities, allows CR transport on much shorter buoyancy
  timescales and solves the CR overpressure problem
  \citep{2008MNRAS.384..251G}.}

This work studies the consequences of this CR-heating model with the aim of
putting forward an observationally well-founded and comprehensive model. In
particular, we intend to eliminate two of the most uncertain assumptions of this
process, the mixing of CRs with the ambient plasma and the CR abundance in the
central cluster regions. Moreover, we tackle the question about thermal balance
in the core region of a CC cluster, but constrained to this model in which CR
Alfv{\'e}n wave heating competes with radiative cooling. To this end, we focus
on the closest active galaxy that is interacting with the cooling cluster gas,
and which provides us with the most detailed view on AGN feedback possible. This
selects M87, the cD galaxy of the Virgo cluster at a distance of $16.7 \pm
0.2$~Mpc \citep{2007ApJ...655..144M}, where $1\arcmin$ corresponds to 4.85
kpc. It is well studied not only at X-rays \citep[e.g.,][]{2010MNRAS.407.2046M}
but also and especially at non-thermal wavelengths, i.e., in the radio
\citep{2000ApJ...543..611O, 2012A&A...547A..56D} and $\gamma$-ray regime
\citep{2012MPLA...2730030R}.

In Section~\ref{sec:NT} we show how new observations of M87 by LOFAR, \Fermi,
and H.E.S.S. can shed light on the injection and mixing of CRs into the ambient
plasma and estimate the CR pressure contribution in the central regions.
Section~\ref{sec:heating} discusses the balance of CR heating with radiative
cooling. Particular attention is paid to local thermal stability of the global
thermal equilibrium that we find.  In Section~\ref{sec:predictions} we present
predictions of our model that include gamma-ray, radio, and Sunyaev--Zel'dovich
(SZ) observations and discuss how those can be used to quantify open aspects of
the presented heating mechanism. Finally, we compare our model with other
solutions to the ``cooling flow problem'' and conclude in Section
\ref{sec:C}. In Appendix~\ref{sec:stability}, we derive a thermal instability
criterion for a fluid that is cooling by thermal bremsstrahlung and line
emission and is heated by a generic mechanism while allowing for thermal and CR
pressure support. In Appendix~\ref{sec:v_A}, we explore the effect of a
density-dependent Alfv{\'e}n speed on our stability considerations, and in
Appendix~\ref{sec:cond}, we compare the efficiency of CR heating to thermal
conduction.

\section{A Non-Thermal Multi-messenger Approach}
\label{sec:NT}

\subsection{Cosmic-ray Injection: Insights by LOFAR} 
\label{sec:LOFAR}

\begin{figure}
\begin{center}
\includegraphics[width=\columnwidth]{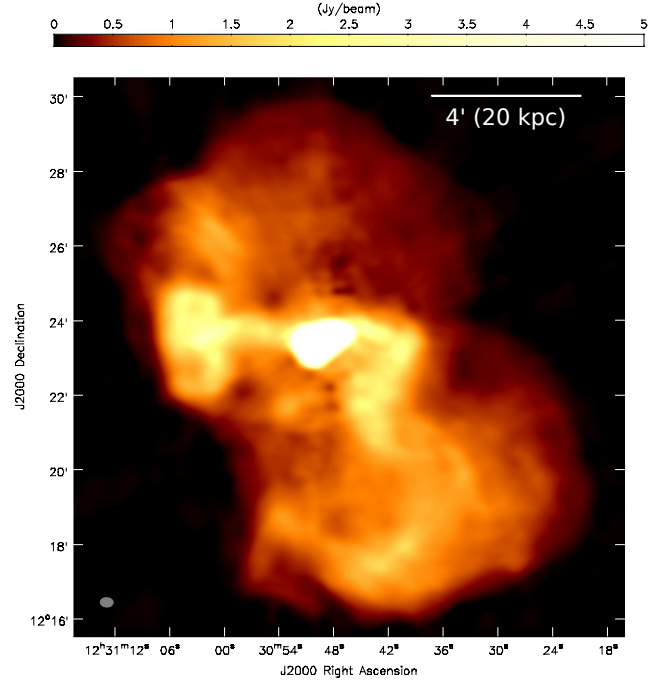}
\end{center}
\caption{LOFAR image of Virgo A at 140 MHz \citep[reproduction with kind
  permission of][]{2012A&A...547A..56D}. The rms noise level is $\sigma =
  20~\rmn{mJy~beam}^{-1}$, the flux peak is $101~\rmn{mJy~beam}^{-1}$ and the
  beam size is $21\arcsec \times 15\arcsec$ (ellipse in the bottom-left
  corner). Note that the radio halo emission fills the entire CC region (in
  projection), indicating that CRs are distributed over $4 \pi$ steradian by
  means of turbulent advection and CR streaming. The comparably sharp halo
  boundary, which coincides with the 1.4 GHz radio emission
  \citep{2000ApJ...543..611O}, suggests a magnetic confinement of CRs to the
  core region.}
\label{fig:virgoA}
\end{figure}

The characteristic frequency $\nu_s$ at which a CR electron (CRe) emits
synchrotron radiation scales with the magnetic field strength $B$ and the CRes'
cooling time $\tau$ as $\nu_s \propto B^{-3} \tau^{-2}$ (assuming the
strong-field limit, $B>3.2\,(1+z)^2\,\mu$G, for which inverse Compton (IC)
cooling is negligible). Hence, observations at lower radio frequencies should
enable us to probe older fossil CRe populations. Those are expected to
accumulate at lower energies, which would imply a low-frequency spectral
steepening. Equivalently, such observations should unveil the time-integrated
non-thermal action of AGN feedback that is expected to be spread out over larger
spatial scales. However, recent LOFAR observations of Virgo A \citep[][see also
Figure~\ref{fig:virgoA}]{2012A&A...547A..56D}, the radio source of M87, provided
two unexpected findings. (1) The observations do not reveal a previously hidden
spectral steepening toward low frequencies and show the extended radio halo to
be well confined within boundaries that are identical at all frequencies
\citep{2012A&A...547A..56D}. The possibility that this may point to a special
time, which is characterized by a recent AGN outburst after long silence, is
unlikely because statistical studies of the radio-mode feedback in complete
samples of clusters imply that the duty cycle of AGN outbursts with the
potential to heat the gas significantly in CC clusters is $\geq60\%$ and could
approach unity after accounting for selection effects
\citep{2012MNRAS.427.3468B}. (2) While the core region displays a radio spectral
index\footnote{We define spectral indices according to
  $F_\nu\propto\nu^{-\alpha_\nu}$.} of $\alpha_\nu=0.6\pm0.02$ (consistent with
an injection CR spectral index of $\alpha=2\alpha_\nu+1=2.2\pm0.04$), the
low-frequency halo spectrum outside the central region (of diameter $\sim2$
arcmin) is considerably steeper with $\alpha_\nu=0.85$. As suggested by
\citet{2012A&A...547A..56D}, the most probable explanation of this steepening at
low frequencies is strong adiabatic expansion in combination with a
superposition of differently synchrotron-aged CRe populations. Adiabatic
expansion across a volume expansion factor $f=V_2/V_1 = \rho_1/\rho_2>1$ from
the compact source to the more dilute halo regions shifts all CR momenta by the
factor $p_2=f^{-1/3}p_1$. CRes that are leaving the lobes have spent different
time intervals in regions of strong magnetic fields, which leads to a
distribution of spectra with different break momenta. Superposition of those
spectra steepens the spectrum down to low frequencies {\em if} the cooling time
of the most strongly cooled CRes has been comparable to the transport time from
the core to the current emitting region.

\begin{figure}
\begin{center}
\includegraphics[width=\columnwidth]{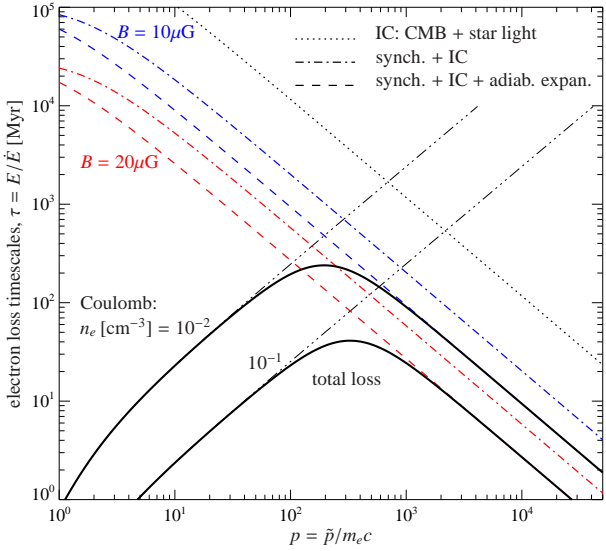}
\end{center}
\caption{Cooling time of electrons for typical conditions in the central region
  of M87 as a function of normalized electron momentum, where $\tilde{p}$ is the
  physical momentum. During their advective transport in the radio lobe, the
  (high-energy) electrons experience inverse Compton (IC), synchrotron, and
  adiabatic losses, for which we assume an expansion factor of $f=10$. After the
  electrons mix with the ambient thermal plasma, Coulomb interactions can
  thermalize them.  Electrons with kinetic energy $E\simeq p\, m_\rmn{e} c^2
  \simeq 150$~MeV are the most long-lived, with a maximum cooling time of
  $\tau\simeq 40$~Myr that corresponds to the radio halo age.}
\label{fig:timescales}
\end{figure}

In Figure~\ref{fig:timescales}, we show the cooling time of CRes \citep[see][for
definitions]{2008MNRAS.385.1211P}. CRes of energy $\sim200~(400)$ MeV that are
responsible for $\sim25~(100)$ MHz radio emission in a $20\mu$G field have
energy loss timescales of $\sim 150~(75)$ Myr, which drops further by the factor
$f^{1/3}$ after accounting for adiabatic expansion. CRes can cool through IC
interactions with various seed photon fields, including the cosmic microwave
background, starlight, and dust-reprocessed infrared emission. Within the
central radio halo region ($r<34$~kpc), the star-formation-induced photon fields
at least contribute as much as the cosmic microwave background
\citep{2011PhRvD..84l3509P}, which motivates our choice of a seed photon energy
density that is twice that of the cosmic microwave background. Synchrotron
cooling in equipartition magnetic fields of $10$--$20\,\mu$G
\citep{2012A&A...547A..56D} and adiabatic expansion substantially decreases the
cooling times further during the advective transport of CRes in the radio
lobes. Those rise in the cluster atmosphere (either buoyantly or driven by the
jet) and uplift dense gas from the central regions
\citep{2010MNRAS.407.2063W}. Those rising motions of light relativistic fluid
cause strong downdrafts at the sides of the lobes, which induce Rayleigh--Taylor
and Kelvin--Helmholtz instabilities that disrupt the lobes and excite turbulence,
which mixes the CRes with the dense uplifted ambient plasma. This picture
appears to be supported by radio maps of M87 \citep{2000ApJ...543..611O,
  2012A&A...547A..56D} that show a diffusely glowing radio halo surrounding the
lobes, which themselves show the characteristic mushroom shape of a
Rayleigh--Taylor instability (see Figure~\ref{fig:virgoA}). This radio map
demonstrates that the radio halo emission fills the entire CC region (in
projection), indicating that CRs are distributed over $4 \pi$ steradian by means
of turbulent advection and CR streaming. The comparably sharp halo boundary,
which exactly coincides with the emission region at 1.4 GHz
\citep{2000ApJ...543..611O}, suggests a magnetic confinement of CRs to the core
region.  Coulomb interactions of the released CRes with the dense ambient ICM
cause them to be thermalized after a maximum cooling time of $\tau\simeq
40$~Myr, which corresponds to the radio halo age (see
Figure~\ref{fig:timescales}). Thus, this naturally explains the absence of a
fossil CRe signature in the low-frequency radio spectra as well as the spatially
similar morphology of Virgo A at all frequencies.

Diffusive shock acceleration operates identically on relativistic particles of
the same rigidity ($R=\tilde{p}c/z e$, where $\tilde{p}$ is the physical
momentum and $z$ is the ion charge), and unlike electrons, protons do
not suffer an injection problem into the acceleration process and have
negligible radiative losses. Hence we expect CR protons (CRps) to dominate the
pressure of the lobes\footnote{In fact, $\alpha$-particles carry a significant
  fraction of the total CR energy, which we absorb into the proton spectrum with
  the following line of reasoning.  A GeV energy $\alpha$-particle can be
  approximated as an ensemble of four individual nucleons traveling together
  due to the relatively weak MeV nuclear binding energies in comparison to the
  kinetic energy of relativistic protons. See \citet{2007A&A...473...41E} for an
  extended discussion.} and to exhibit a similar injection spectrum as the CRes
in the central regions ($\alpha\approx 2.2$). Similarly to the CRes, CRps should
also be released from the lobes and mixed into the ambient ICM. Since
adiabatic expansion leaves their spectral shape invariant and CRps have
negligible radiative losses, we do not expect a spectral steepening of the CRp
population in the halo. A necessary consequence of this picture are hadronic
interactions of CRps with the ambient gas. This should produce a steady
gamma-ray signal from M87 \citep{2003A&A...407L..73P}, with a spectral index
above GeV energies of $\alpha_\gamma=\alpha\approx 2.2$.

\subsection{Cosmic-ray Pressure Estimates from \Fermi and H.E.S.S.} 
\label{sec:Fermi}

To test this hypothesis, we turn to the gamma-ray regime. A joined unbinned
likelihood fit with a power law ($dN/dE\propto E^{-\alpha_\gamma}$) to the
\Fermi and H.E.S.S. data of M87's low state yields
$F(>1~\rmn{GeV})=1.35(\pm0.35)\times10^{-9}\,\rmn{photons~cm}^{-2}\,\rmn{s}^{-1}$
and $\alpha_\gamma=2.26\pm0.13$ \citep{2009ApJ...707...55A}, which is consistent
with our expectation if the gamma rays are produced in hadronic CRp--p
interactions. Moreover, a comparison of the \Fermi light curve for the last 14
months with the initial 10 month data revealed no evidence of variability in
the flux above 100 MeV \citep{2012ApJ...746..151A,
  2009ApJ...707...55A}.\footnote{A light-curve analysis of 4 yr of \Fermi data
  does not show any sign of variability either (F.{\ }Rieger, private
  communication).}  Gamma-ray upper limits on M87 set by EGRET
\citep{2003ApJ...588..155R} are consistent with the steady flux seen by
\Fermi. This demonstrates that the EGRET and \Fermi observations of M87, which
together span more than two decades, show no apparent increase in
flux. Additionally, the {\em low flux state} at very high energies $E>100$~GeV
does not provide any hint of variability \citep{2006Sci...314.1424A}, although
the low signal statistics renders such a test very difficult. Encouraged by the
spectral similarity of radio and gamma-ray observations and the constant
gamma-ray flux, we will now assume that the gamma rays are tracing hadronic CR
interactions.

To estimate the CR pressure that is implied by the observed gamma-ray emission,
we choose a constant CR-to-thermal pressure ratio, $X_\CR=P_\CR/P_\rmn{th}$
(which we require for self-consistency, as we will see in
Section~\ref{sec:heating}).  We assume an isotropic one-dimensional CR
distribution function in momentum space,\footnote{The three-dimensional
  distribution function is $f^{(3)}(p)=f(p)/(4\pi\,p^2)$.}
\begin{equation}
  \label{eq:fCR}
  \frac{d^4N}{dp\,d^3x}=C p^{-\alpha}\theta(p-q), 
\end{equation}
where $\theta$ denotes the Heaviside step function, $p=\bar{p}/mc$ and $q$
denote the normalized momentum and low-momentum cutoff, $C$ denotes the
normalization, and $\alpha=2.26$. We decided in favor of the spectral index of
the gamma-ray emission, which likely probes the prevailing CRp distribution in
the radio halo region. In contrast, the radio emission of the central region may
signal current acceleration conditions that, albeit generally very similar, may
be slightly different from the ones that have accelerated the CRs in the past,
which have now been released.

We use a fit to the spherically symmetric thermal pressure profile as inferred
from X-ray observations \citep{2002A&A...386...77M, 1994Natur.368..828B}.  The
thermal pressure is $P_\rmn{th} = n kT = n_e kT /\mu_e$, where $\mu_e = n_e/n =
\mu\, (X_\rmn{H}+1)/2 \simeq 0.517$ for a fully ionized gas of primordial
element composition with a hydrogen mass fraction $X_\rmn{H}=0.76$ and mean
molecular weight $\mu = 0.588$. The electron density and temperature
profiles\footnote{In Virgo, the gas is observed to have a maximum temperature
  $T\simeq 3\times10^7~$K at a cluster-centric radius of $\sim200~$kpc
  \citep[corresponding to $40\arcmin$;][]{1994Natur.368..828B}.} of the central
region of the Virgo cluster ($r\aplt100~\rmn{kpc}$ corresponding to $20\arcmin$)
are given by fits to the X-ray profiles and have the form
\begin{eqnarray}
\label{eq:ne}
n_e (r) &=& 
\left[
  \sum_{i=1}^2 n_i^2 \, \left( 1+\frac{r^2}{r^2_i}
  \right)^{-3 \beta_i} 
\right]^{1/2},\\
\label{eq:Te}
T_e (r) &=& T_0 + (T_1 - T_0)\,
\left[ 1+\left( \frac{r}{r_t} \right)^{-1}\right]^{-1},
\end{eqnarray}
where $(n_1,n_2) = (0.15,0.013)~\rmn{cm}^{-3}$, $(r_1,r_2) =
(1.6,20)~\rmn{kpc}$, $(\beta_1,\beta_2) = (0.42,0.47)$, $(kT_0,kT_1) =
(1,3)~\rmn{keV}$, and $r_t = 13.5~\rmn{kpc}$. 

The sharp drop of radio morphology of Virgo A's halo suggests that the CRps are
also confined to a central region of radius $r_\rmn{max}=34$~kpc. Hence, we
integrate the gamma-ray flux due to hadronic CRp--p interactions from within a
spherical region bounded by $r_\rmn{max}$. We use a semi-analytic formalism
\citep{2004A&A...413...17P} that has been modified to constrain the CR-to-thermal
{\em pressure} ratio \citep{2008MNRAS.385.1211P}. We obtain $X_\CR=0.38$ for a
vanishing low-momentum cutoff, $q=0$. Accounting for a non-zero low-momentum
cutoff of the CRp distribution as a result of Coulomb cooling, we obtain
$X_\CR=0.31~(0.34)$ for $q=0.58~(0.25)$, which corresponds to a Coulomb cooling
time of CRps of $\tau=40$~Myr, assuming an ambient electron density of
$n_e=10^{-1}\,\left(10^{-2}\right)~\rmn{cm}^{-3}$. Since there is considerable
uncertainty about the shape of the low-momentum regime of the CR distribution
function, we decided to adopt the most conservative value of $X_\CR=0.31$ for
the remainder of this work. We note that a fresh supply of CRps could lower $q$
and therefore increase the CR pressure.

We checked that the radio emission from secondary CRes, which are inevitably
injected by hadronic CRp--p reactions, has a subdominant contribution to the
observed radio emission. While the total radio emission due to secondaries is
well below the observed flux density of the radio halo of M87, future
high-frequency observations at $\apgt10$~GHz of the external halo parts may be
able to probe the hadronically induced radio emission (see
Section~\ref{sec:radio} for more details). At these frequencies, the
hadronically induced emission component cannot any more be neglected and
complicates inferences on the CRe pitch-angle distribution in different spectral
aging models of radio-emitting CRes.

In general, cluster merger and accretion shocks are also expected to accelerate
CRs, which add to the CR population injected by the central
AGN. Non-observations of gamma-ray emission at GeV and TeV energies limit their
contribution to $X_\CR<0.017$ in the Coma and Perseus clusters
\citep{2012ApJ...757..123A, 2012A&A...541A..99A} and typically to less than a
few per cent for the next best targets \citep{2010ApJ...717L..71A,
  2011PhRvD..84l3509P}. A joint likelihood analysis searching for spatially
extended gamma-ray emission at the locations of 50 galaxy clusters in 4 yr of
\Fermi-LAT data does not reveal any CR signal and limits the CR-to-thermal
pressure ratio to be below 0.012--0.014 depending on the morphological
classification \citep{2013arXiv1308.5654T}. We hence neglect their contribution
in the center, where the AGN-injected population apparently dominates.

Our findings for $X_\CR$ are in agreement with constraints on the non-thermal
pressure contribution in M87 that have been derived by comparing the
gravitational potentials inferred from stellar and globular cluster kinematics
and from assuming hydrostatic equilibrium of the X-ray emitting gas
\citep{2010MNRAS.404.1165C}. Depending on the adopted estimator of the optical
velocity dispersion, these authors find a non-thermal pressure bias of
$X_\rmn{nt}=P_\rmn{nt}/P_\rmn{th}\approx0.21$--0.29 (which is even higher in M87
for more sophisticated estimates that are less sensitive to the unknown orbit
anisotropy). There are several effects that could (moderately) increase
$X_\rmn{nt}$. (1) A more extended distribution of $P_\rmn{nt}$
in comparison to $P_\rmn{th}$ could hide a larger non-thermal pressure
contribution (since only the pressure gradient enters into the equation for
hydrostatic equilibrium), (2) the presence of an X-ray ``invisible'' cool gas
phase that is coupled to the hot phase could bias $X_\rmn{nt}$ low
\citep{2008MNRAS.388.1062C}, and (3) the tension in the orbit anisotropy
estimates of the stars and globular clusters may call for triaxial potential
models \citep{2001ApJ...553..722R}.

\subsection{Pressure Content of Radio Lobes}
\label{sec:content}

So far, we implicitly assumed that the pressure of radio lobes is provided by a
relativistic component. While it is unknown what physical component provides the
dominant energetic contribution to radio lobes, here we will review the reasons
for our assumption.  Equipartition arguments for the radio-emitting lobes show
that the sum of CRes and magnetic fields can only account for a pressure
fraction of $\simeq10\%$ in comparison to the external pressure, with which the
lobes are apparently in approximate hydrostatic equilibrium
\citep{2003ApJ...585..227B, 2012A&A...547A..56D}. However, these estimates rely
on power-law extrapolations from the high-energy part of the CRe distribution
(with Lorentz factor $\gamma\apgt 10^3$, which was visible so far at radio
frequencies $\nu\apgt 300$~MHz) to lower energies that dominate the CRe energy
budget, leaving the possibility of an aged fossil CRe population that would
dominate the pressure. LOFAR observations have now unambiguously demonstrated
the absence of such a hypothetical fossil CRe population in M87
\citep{2012A&A...547A..56D}. Since CRps should be accelerated at least as
efficiently as the visible CRes and since CRps have negligible radiative cooling
rates (in comparison to CRes), we make the very plausible assumption in this
work that CRps provide the dominating pressure component of lobes (and work out
a few smoking-gun tests for this scenario in
Section~\ref{sec:predictions}). Such a CRp population could originate from
entrained material as a result of the interaction of a Poynting-flux-dominated
jet with the surrounding plasma. This entrained material can get accelerated by
processes such as internal shocks or Kelvin--Helmholtz turbulence to give rise to
a spine-layer jet geometry at larger scales that is needed to explain the
multi-frequency data of some AGNs \citep[e.g., in
NGC 1275;][]{2010ApJ...710..634A}.

The alternative scenario of a hot but subrelativistic (Maxwellian) proton
population is unlikely because of three reasons. (1) CRp Coulomb cooling
timescales are too short \citep[see, e.g.,][]{2007A&A...473...41E}, implying a
burst-like heating once the content has been set free as a result of disrupting
the radio lobes by the action of Kelvin--Helmholtz and Rayleigh--Taylor
instabilities. However, there is no signature in the X-ray observations of such
a localized heating around the disrupted lobes \citep{2010MNRAS.407.2046M}. (2)
Moreover, LOFAR observations show no sign of such a $\sim100$~MeV electron
distribution, requiring a helicity fine tuning for the acceleration process. (3)
Adiabatic expansion of a composite of relativistic and non-relativistic
(thermal) populations favors the CR over the non-relativistic population since
$P_\CR / P_\rmn{th} \propto f^{1/3}$, due to the softer equation of state of
CRs. Here we assumed an ultra-relativistic CR population with an adiabatic
exponent of $\gamma=4/3$ and $f=V_2/V_1 = \rho_1/\rho_2$ is the volume expansion
factor starting from the densities at which the two populations got energized.

\section{Cosmic-Ray Heating versus Radiative Cooling}
\label{sec:heating}

The previous arguments provide strong evidence that CRs are released from the
radio lobes of M87 and mixed into the ICM on a timescale of $\tau \simeq
40$~Myr. Interpreting the observed gamma-ray emission as a signature of hadronic
CRp--p interactions enabled us to estimate the CR pressure contribution. We now
turn to the question whether the resulting CR Alfv\'en wave heating is powerful
enough to balance radiative cooling and what this may imply for thermal
stability of the gas in M87 and CCs in general.

\subsection{Cosmic-ray Heating}
\label{sec:CR}

Advection of CRs by the turbulent gas motions produces centrally enhanced
profiles. In contrast, CR streaming along magnetic field lines implies a net
outward flux of CRs that causes the profiles to flatten
\citep{2011A&A...527A..99E}.  A necessary requirement for CR streaming to work
is a centrally peaked profile of the CR number density.  Do we have any evidence
for this in M87? Radio maps of Virgo A \citep{2000ApJ...543..611O,
  2012A&A...547A..56D} suggest that turbulent advection, driven by the rising
lobes, is strong enough to maintain a sufficiently strong inward transport of
CRs. Alternatively, the smoothly glowing radio halo that fills the central
region for $r<r_\rmn{max}$ may also be caused by escaping CRs from the lobes
already at the early stages of their rise. Following observational and
theoretical arguments, the current distribution of the CR pressure appears to
follow a centrally peaked profile.

The CR Alfv{\'e}n wave heating term is given by \citep{1971ApJ...163..503W,
  2008MNRAS.384..251G}
\begin{equation}
  \label{eq:CR}
  \mathcal{H}_\CR = -\bvel_A \bcdot \boldsymbol{\nabla} P_\CR
  = -\bvel_A \bcdot \left(\mathbfit{e}_r\nabla_r \bra P_\CR\ket_\Omega + 
    \mathbfit{e}_l\frac{\delta P_\CR}{\delta l}\right),
\end{equation}
where $\bvel_A = \mathbfit{B}/\sqrt{4 \pi \rho}$ is the Alfv{\'e}n velocity,
$\rho$ is the gas mass density, and we decompose the local CR pressure gradient
into the radial gradient of the azimuthally averaged CR pressure profile, $\bra
P_\CR\ket_\Omega$, and a fluctuating term, $\mathbfit{e}_l \delta P_\CR/\delta l
\equiv \boldsymbol{\nabla} P_\CR - \mathbfit{e}_r \nabla_r \bra
P_\CR\ket_\Omega$, which in general also points in a non-radial direction,
defined by the unit vector $\mathbfit{e}_l$. In steady state, CR streaming
transfers an amount of energy to the gas per unit volume of
\begin{equation}
  \label{eq:heat}
  \Delta\eps_\rmn{th} = -\tau_A \bvel_A \bcdot \boldsymbol{\nabla} P_\CR 
  \approx P_\CR = X_\CR P_\rmn{th},
\end{equation}
where we used the Alfv{\'e}n crossing time $\tau_A = \delta l/\vel_A$ over the
CR pressure gradient length, $\delta l$. Comparing the first and last term of
this equation suggests that a constant CR-to-thermal pressure ratio is a
necessary condition if CR streaming is the dominant heating process, i.e., the
thermal pressure profile adjusts to that of the streaming CRs in equilibrium so
that we will take the radial profile of $P_\rmn{th}$ as a proxy for $P_\CR$.

The observed radial profile of $P_\rmn{th}$ averages over the local pressure
inhomogeneities within a given radial shell, especially toward the center where
the azimuthally averaged radial profile plateaus out. However, such a constant
$P_\rmn{th}$ at the center is in strong disagreement with the fluctuating
thermal pressure maps \citep[see Figure 7 of][]{2010MNRAS.407.2046M}. Those
pressure fluctuations could have their origin in shocks, turbulence,
perturbations of the gravitational potential, or entropy variations of advected
gas that has not had time to come into pressure equilibrium with the
surroundings. To illustrate one of these possibilities, we consider pressure
fluctuations due to weak shocks and expand the Rankine--Hugoniot jump conditions
in the limit of small Mach number, $\M_1 - 1 = \eps\ll 1$:
\begin{equation}
  \label{eq:deltaP}
  \frac{P_2}{P_1} = \frac{2 \gamma_\rmn{th} \M_1^2 - (\gamma_\rmn{th}-1)}{\gamma_\rmn{th}+1} \approx
  1 + \frac{4 \gamma_\rmn{th}}{\gamma_\rmn{th}+1}\,\eps = 1 + \frac{5}{2}\,\eps,
\end{equation}
where subscripts 1 and 2 denote upstream and downstream quantities in the shock
rest frame and we assume $\gamma_\rmn{th}=5/3$ in the third step. Adopting a constant
CR-to-thermal pressure ratio, we obtain an estimate for the CR heating due to
pressure fluctuations as a result of a sequence of weak shocks, i.e., the second
term on the right-hand side of Equation~\eqref{eq:CR},
\begin{equation}
  \label{eq:delP}
  \frac{\delta P_\CR}{\delta l} \approx 
  -\frac{5}{2}\, \frac{\eps\, X_\CR P_\rmn{th}}{r}.
\end{equation}
Here we assume that the characteristic size scale of the pressure fluctuations
is of order the cluster-centric radius, which can be realized, e.g., for
spherically symmetric shocks whose curvature radius increases as $r$. We note
that some of the other mechanisms that may be responsible for the observed
pressure fluctuations can in principle have a different scaling with radius and
hence have a modified radial dependence of the fluctuating term in
$\mathcal{H}_\CR$, in particular at the center. It will be subject to future
work to quantify the contribution of those processes to the CR heating rate.

We choose conservative values for the magnetic field strengths that are about
half as much as the minimum energy estimates presented in
\citet{2012A&A...547A..56D}. We adopt $B = 20\,\mu\rmn{G}\,\sqrt{n_e / 0.1
  \rmn{cm}^{-3}}$, which yields $B = 24~(5)~\mu$G for the central value (at the
halo boundary, $r_\rmn{max} = 34$~kpc). The density scaling yields a constant
Alfv{\'e}n velocity $\vel_A \approx 130~\rmn{km~s}^{-1}$. The plasma beta factor
$\beta=P_\rmn{th}/P_B$ increases from 20 (at the center) to 47 (at
$r_\rmn{max}$), which corresponds to a magnetic pressure contribution of $X_B =
0.05$ (center) and 0.02 ($r_\rmn{max}$).
\vspace{2em}

\subsection{Global Thermal Equilibrium}
\label{sec:global}

Radiative cooling of cluster gas is dominated by thermal bremsstrahlung and
collisional excitation of heavy ions to metastable states and the subsequent
line emission following de-excitation (the latter of which dominates for
particle energies $\aplt3$~keV). These processes are characterized by the
cooling rate
\begin{equation}
  \label{eq:cool}
  \mathcal{C}_\rmn{rad} = n_e n_t \Lambda_\rmn{cool}(T, Z), 
\end{equation}
where the target density $n_t$ is the sum of ion densities and we defined the
cooling function $\Lambda_\rmn{cool}(T, Z)$ that depends on temperature $T$ and
metal abundance $Z$. The cooling function is fairly flat between the range of
temperatures of $kT_0=1$--$2~\rmn{keV}$, we encounter in the central regions,
bounded by the radio halo. The X-ray inferred metallicity map shows substantial
spatial variations on small scales ranging between $Z_0 = 0.7$ and
$1.3~Z_\odot$, which are of similar amplitude as the overall radial trend in $Z$
\citep{2010MNRAS.407.2046M}. Hence, we decided to bracket the cooling function
for these ranges of temperature and metallicity, which yields
$\Lambda_\rmn{cool} (T_0, Z_0) \approx (1.8$--$2.8)\times 10^{-23}\,
\rmn{erg~cm}^3~\rmn{s}^{-1}$ while assuming collisional ionization equilibrium
\citep[see Figure~8 of][]{1993ApJS...88..253S}.

\begin{figure}
\begin{center}
\includegraphics[width=\columnwidth]{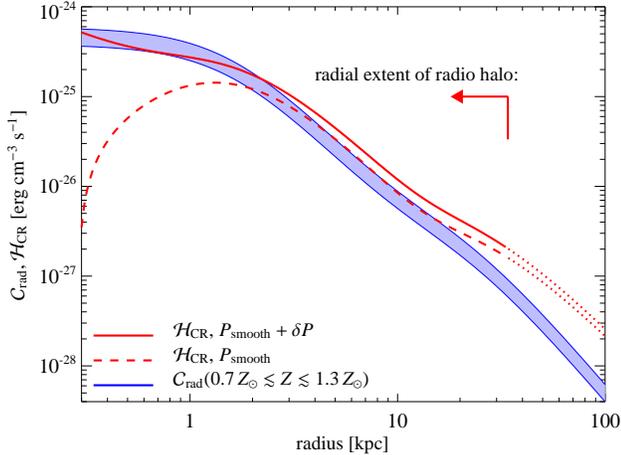}
\end{center}
\caption{CR heating vs. radiative cooling rates. The cooling rates (blue)
  account for a fluctuating metal distribution in the range 0.7--$1.3\,Z_\odot$
  and are globally balanced at each radius by CR Alfv{\'e}n wave heating. We show
  the CR heating rate $\mathcal{H}_\CR$ due to the azimuthally averaged pressure
  profile that ``smoothes'' out the CR pressure gradient (dashed red) and the
  $\mathcal{H}_\CR$ profile, where we additionally account for pressure
  fluctuations due to weak shocks of typical Mach number $\M=1.1$ (solid
  red). The CR heating rates become very uncertain outside the boundary of the
  radio halo of Virgo A (dotted red).}
\label{fig:heat}
\end{figure}

Figure~\ref{fig:heat} shows the radial profile of the radiative cooling rate
$\mathcal{C}_\rmn{rad}$ and that of the CR heating rate $\mathcal{H}_\CR$
assuming $\vel_A=\rmn{const}$. For radii $r\apgt2$~kpc, CR heating balances
cooling already for the ``smoothed'' gradient pressure term ($\nabla_r \bra
P_\CR\ket_\Omega$). The central pressure plateau causes the CR pressure gradient
and hence the CR heating to drop. To model the large visible fluctuations in
$P_\rmn{th}$ (and hence in $P_\CR$ according to Equation \eqref{eq:heat}), we
assume that they are caused by weak shocks of typical Mach number $\M=1.1$ with
a size scale that is of order the radius (i.e., we adopt $\eps=0.1$ in
Equation~\eqref{eq:deltaP}). This fluctuating pressure term appears to be critical
in balancing the cooling rate with CR heating in the central region for
$r<2$~kpc. We conclude that CR Alfv{\'e}n wave heating can globally balance
cooling, hence justifying our assumption of a constant CR-to-thermal pressure.
In Appendix~\ref{sec:v_A}, we show that a varying $\vel_A$ (within physical
bounds) does not change our main conclusion about the ability of CR heating to
balance radiative cooling globally although there exists the possibility that
heating dominates cooling at the external halo regions, which would cause these
to adiabatically expand.

Particularly notable is the non-trivial result that the radial shapes of
$\mathcal{C}_\rmn{rad}\propto n^2$ and $\mathcal{H}_\CR\propto \nabla_r
P_\rmn{th}$ closely track each other.  In fact, if CR heating dominates,
$P_\rmn{th}$ should be determined by the pressure profile of the streaming CRs,
and hence this resemblance of the radial heating and cooling rate profiles is an
important signature of self-consistency. The sharp boundaries of the radio halo
of Virgo A may suggest that CRs are confined to within its boundaries, perhaps
as a result of azimuthally wrapped magnetic fields. Hence, the CR heating rates
become very uncertain outside the maximum halo radius.

We will argue that this similarity of the CR heating and radiative cooling
profiles is not a coincidence in M87, but rather generically expected in models
of CR Alfv{\'e}n wave heating due to the presence of a physical attractor
solution for non-equilibrium initial conditions (albeit constrained to the time
average). Global equilibrium requires matching heating and cooling rates at any
radius, i.e., similar radial profiles as well as a matching overall
normalization. The latter is a consequence of the self-regulating property of
AGN feedback. If $\mathcal{H}_\CR<\mathcal{C}_\rmn{rad}$ at some small radius,
this triggers runaway cooling and eventually accretion onto the central black
hole. Phenomenologically, this is connected to the launch of relativistic jets
\citep{2007ARA&A..45..117M} that carry a fraction of the accretion energy to kpc
scales and inflate radio lobes with CRs (see Section~\ref{sec:content}) that
boost the CR heating rate to become larger or equal to the radiative cooling
rate \citep{2008MNRAS.384..251G}. If $\mathcal{H}_\CR> \mathcal{C}_\rmn{rad}$,
the CR heating overpressurizes this region on the Alfv{\'e}nic crossing
time. This heated patch cannot radiate away the thermal energy and expands
quasi-adiabatically in the cluster atmosphere until $\mathcal{H}_\CR\simeq
\mathcal{C}_\rmn{rad}$ (since we assume here that the Alfv{\'e}n velocity is
smaller than the sound speed, see Section~\ref{sec:CR}). This implies a transfer
of gas mass to larger radii and a flattening of the density profile.

To demonstrate that the radial heating and cooling profiles generically match in
this model, we use a constant gas-to-dark matter ratio inside the dark-matter
scale radius, i.e., $n\propto r^{-1}$, which is characteristic for the central
density profile in CCs (and which applies to the density profile of the Virgo
cluster for $2~\rmn{kpc}\aplt r\aplt 100$~kpc, see
Equation~\eqref{eq:ne}). Outside the temperature floor, we adopt a temperature
profile with a small positive slope, $T\propto r^{\alpha_t}$ (with
$\alpha_t\aplt 0.3$ typically), so that the pressure profile scales as
$P_\rmn{th}=n kT\propto r^{\alpha_t-1}$. As a result, we obtain a scaling of the
cooling and CR heating rates with radius of $\mathcal{C}_\rmn{rad} \propto
r^{-2}$ and $\mathcal{H}_\CR\propto d (r^{\alpha_t-1})/dr \propto
r^{\alpha_t-2}$. Those are identical for a constant temperature
($\alpha_t=0$). For a smoothly rising temperature profile ($\alpha_t\aplt 0.3$)
the heating rate profile is tilted and slightly favored over the cooling rates
at larger radii, provided they match up at some inner radius (see
Figure~\ref{fig:heat}). A corollary of this consideration is that the onset of
cooling is smoothly modulated from the outside in, as we will see in the
following.

\subsection{Local Stability Analysis}
\label{sec:local_stability}

CR heating appears to be able to keep the gas in {\em global} thermal
equilibrium in M87. However, this does not necessarily imply {\em local} thermal
stability. We consider small isobaric perturbations to this equilibrium
configuration, which maintain pressure equilibrium at each radius. A necessary
condition for maintaining hydrostatic equilibrium is a short dynamical time
$\tau_\rmn{dyn}$ in comparison to the CR heating time $\tau_A$.  We obtain the
following condition on the dynamical timescale
\begin{equation}
  \label{eq:dyn}
  \tau_\rmn{dyn} \simeq \sqrt{\frac{R}{g}} \approx \frac{R}{c_s} \stackrel{!} {<}
  \frac{R}{\vel_A} \approx \tau_A,
\end{equation}
where $c_s$ is the sound speed and $g = GM(<R)/R^2$ is the gravitational
acceleration. This inequality can be rewritten to yield $c_s / \vel_A \simeq
\sqrt{\beta\gamma_\rmn{th}/2} > 1$, where $\beta$ is the plasma beta factor and
$\gamma_\rmn{th}=5/3$. For magnetic field estimates adopted here, we obtain $c_s
/ \vel_A \simeq 4$ (6.3) at the center (at the halo boundary, $r = 34$~kpc) so
that this inequality is always fulfilled.  

In Appendix~\ref{sec:stability} we derive a general instability criterion for a
fluid that is cooling by thermal bremsstrahlung and line emission and heated by
a generic mechanism with a power-law dependence on $\rho$ and $T$ while allowing
for thermal and CR pressure support. We will start the discussion with a
transparent order-of-magnitude calculation that provides insight into the
physics and complements the formal treatment in Appendix~\ref{sec:stability}.
The generalized Field criterion \citep{1965ApJ...142..531F, 1986ApJ...303L..79B}
for thermal instability in the presence of isobaric perturbations is given by
\begin{equation}
  \label{eq:Field}
  \left.\frac{\partial (\calL / T)}{\partial T}\right|_P < 0,
\end{equation}
where $\calL$ is defined as the net loss function per unit mass such that
$\rho\calL=\mathcal{C}-\mathcal{H}$, where $\mathcal{C}$ and $\mathcal{H}$ are
the cooling and heating rates per unit volume. 

Here, we assume the existence of small-scale tangled magnetic fields that result,
e.g., from magnetohydrodynamic turbulence. A cooling gas parcel necessarily
pulls in the flux-frozen magnetic fields alongside the contracting gas. The CRs
are bound to stay on a given flux tube and will experience adiabatic compression
during the onset of collapse. The streaming motion of CRs into the increasingly
denser gas parcel is driven by the free energy stored in the large-scale CR
pressure gradient. On the Alfv{\'e}nic crossing timescale that is of relevance
for our stability analysis, the CR population does not suffer non-adiabatic
losses such as hadronic interactions \citep[e.g.,][]{2007A&A...473...41E}. In
particular, CR streaming is an adiabatic process for the CR population
\citep{Kulsrud}, implying locally $P_\CR\propto \rho^{\gamma}$, where $\gamma$
is the adiabatic exponent of the CR population.  We emphasize that this {\em
  local} adiabatic behavior of $P_\CR$ does not hold for the {\em equilibrium}
distribution of CRs that is subject to non-adiabatic gain and loss processes,
which modify the ``CR entropy'' (i.e., its probability density of the
microstates).\footnote{This is similar to thermal cluster gas that increases its
  pressure $P_\rmn{th}\propto\rho^{5/3}$ upon adiabatic compression. However,
  non-adiabatic processes such as shocks triggered by gravitational infall,
  non-gravitational heating, and radiative cooling modify the gas entropy. As a
  result, the gas obeys an effective adiabatic exponent
  $\gamma_\rmn{eff}\equiv\partial\ln P_\rmn{th}/\partial \ln \rho \approx
  1.1$--$1.3$ within a cluster that only slowly increases beyond the radii of
  accretion shocks toward the adiabatic value
  \citep[e.g.,][]{2010ApJ...725...91B, 2012MNRAS.422..686C}.}  We obtain the
scaling of the CR heating rate with temperature,
\begin{equation}
\label{eq:HCR_OoM}
\mathcal{H}_\CR = - \bvel_A\bcdot\boldsymbol{\nabla} P_\CR\propto \rho^{\gamma+1/3} 
\propto T^{-\gamma-1/3}.
\end{equation}
Here we identified the characteristic size of our unstable clump with an
exponential pressure scale height $h$ so that the local gradient length scales
as $\delta l \sim h \propto\rho^{-1/3}$, and we assumed $\vel_A=\rmn{const}$. We
suppress the dependence on pressure, which is constant for isobaric
perturbations. The radiative cooling rate scales with temperature as
\begin{equation}
\label{eq:C_OoM}
 \mathcal{C}_\rmn{rad}
  = \rho^2 \left[\Lambda_0 Z^2 T^{1/2} + \Lambda_l(T,Z)\right]
  \propto T^{-3/2} + \eta T^{-2+\chi},
\end{equation}
where the first and second term in the brackets denote the bremsstrahlung and
line-emission cooling. Here we introduce the logarithmic temperature slope of
the line cooling function $\chi$ and the ratio of line-to-bremsstrahlung cooling
rate $\eta$. For an ultra-relativistic CR population with $\gamma=4/3$, we find
a stable fixed point in the bremsstrahlung regime since the slope of the heating
rate ($=-5/3$) is steeper than that of the cooling rate ($=-3/2$). Increasing
(decreasing) the temperature of a perturbed parcel of gas implies a stronger
cooling (heating) that brings the parcel back to the background
temperature. However, in the line-cooling regime (at $(1$--$3)\times10^7$~K), which
is characterized by $\eta\apgt1$ and $\chi\aplt0$, the slope of the cooling rate
($\aplt -2$) is steeper than that of the CR heating rate ($=-5/3$), implying
instability.

\begin{figure}
\begin{center}
\includegraphics[width=\columnwidth]{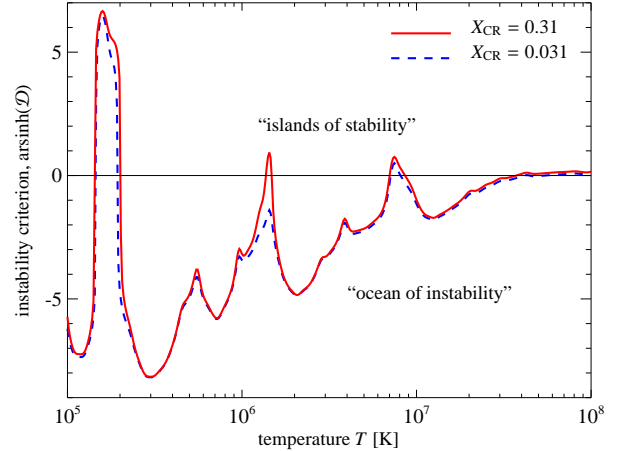}
\end{center}
\caption{Local stability analysis of a fluid for which CR heating globally
  balances radiative cooling by bremsstrahlung and line emission (assuming solar
  metallicity). We show the instability criterion,
  $\mathcal{D}\propto\left.\partial (\calL/T)/\partial T\right|_P$, as a
  function of temperature for parameters appropriate for M87 (red) and for a 10
  times lower CR pressure (blue dashed). Precisely, we show
  $\rmn{arsinh}(\mathcal{D})$, which has a linear (logarithmic) scaling for
  small (large) values of the argument.  Once the gas temperature drops below
  $3\times10^7$~K, it becomes thermally unstable. Assuming that the gas cools
  while maintaining approximate thermal equilibrium, its collapse is halted at
  around $10^7$~K by the first ``island of stability'', which is consistent with
  the temperature floor seen in X-ray observations. However, the high-density
  tail of density fluctuations should be able to cross this island and become
  again subject to thermal instability, likely sourcing some of the observable
  multi-phase gas.}
\label{fig:instability}
\end{figure}

This simple order-of-magnitude calculation is substantiated in
Figure~\ref{fig:instability}, which shows the instability criterion derived in
Equation~\eqref{eq:instability1} and restricted to CR Alfv{\'e}n wave heating as
the sole source of heat in the central region of the Virgo cluster. A gas in
thermal equilibrium ($\rho\calL=0$) is thermally unstable if
\begin{equation}
  \label{eq:instability_CR}
  \mathcal{D} \equiv \frac{1}{2} + \eta\chi
  - \frac{(5/3 - \gamma) (1 + \eta )}{1+X_\CR\gamma} < 0.
\end{equation}
We compute the logarithmic temperature slope of the line cooling function
$\chi(T)$ and the ratio of line-to-bremsstrahlung cooling rate $\eta(T)$ for a
gas at solar metallicity and assume collisional ionization equilibrium
\citep{1993ApJS...88..253S}.  In Figure~\ref{fig:instability}, we show
$\mathcal{D}$ as a function of temperature and adopt values appropriate for M87
($\gamma=1.37$ and $X_\CR=0.31$, see Appendix~\ref{sec:CRA}). For most of the
temperature range, we find that $\mathcal{D}<0$ and the gas is subject to
thermal instability. However, there are a few ``islands of stability'' where
$\mathcal{D}>0$.

The location of such an ``island of stability'' corresponds to a sufficiently
positive slope of the cooling function $\Lambda_\rmn{cool}$ as a function of $T$
(i.e., it is located at the low-temperature side of every line-cooling complex).
Increasing the metallicity $Z$ increases the cooling rate and sharpens the line
complexes. This implies more stable ``islands of stability,'' i.e., they
protrude more above the ``ocean of instability,'' but they also extend over a
slightly smaller temperature range. However, increasing $Z$ does not
significantly change their localization in temperature (as long as the CR
heating can compensate the increasing cooling rate that accompanies the elevated
$Z$).

Apart from parameters associated with radiative cooling (e.g., metallicity),
$\mathcal{D}$ also depends on cluster-immanent parameters associated with the CR
distribution and magnetic fields, i.e., the CR adiabatic exponent $\gamma$, the
CR-to-thermal pressure ratio $X_\CR$, and $\vel_A$. Of those, only $X_\CR$ and
the possible density dependence of $\vel_A$ can in principle vary substantially
among clusters. To address the robustness of the shape of the instability
criterion $\mathcal{D}$ with respect to variations in $X_\CR$, in
Figure~\ref{fig:instability} we additionally show $\mathcal{D}$ for a 10 times
lower CR pressure (which would, however, not suffice to maintain global thermal
equilibrium in M87).  Most importantly, the shape of $\mathcal{D}$ versus
temperature barely changes, except that the second ``island of stability''
vanishes for low values of $X_\CR$. In Appendix~\ref{sec:v_A}, we address how a
varying $\vel_A$ impacts the local stability analysis. In particular, for the
relation $\vel_A\propto\rho^{1/2}$, which represents the extreme case of
collapse solely perpendicular to $\mathbfit{B}$, we find more prominent
``islands of stability'' while still obeying global thermal equilibrium (within
the range of uncertainty).

\subsection{A Critical Length Scale for the Instability}
\label{sec:lambda_crit}

Before we can discuss the fate of the cooling gas, we will be considering the
typical length scale of a region that becomes subject to thermal instability.
We expect cooling gas parcels only to appear in systems whose dimension is
greater than a critical length scale $\lambda_\rmn{crit}$, below which CR
Alfv{\'e}n wave heating smoothes out temperature inhomogeneities. We will
estimate $\lambda_\rmn{crit}$ by considering thermal balance for a cool parcel
of radius $r$ embedded in a medium with an equilibrium CR pressure $P_\CR=X_\CR
P_\rmn{th}$. CR streaming transfers an amount of energy to a given parcel at a
rate $\sim r^3 f_s \vel_A (P_\CR/r)$, where $f_s$ is a suppression factor that
depends on the magnetic topology connecting the parcel with its
surroundings. Line emission and bremsstrahlung cooling can radiate energy from
the cloud at a rate $\sim r^3 n_e n_t \Lambda_\rmn{cool}(Z,T)$. Cooling and CR
heating are thus approximately balancing each other for systems with a radius
\begin{equation}
  \label{eq:lambda_crit}
  \lambda_\rmn{crit} = \frac{f_s \vel_A P_\CR}{n_e n_t \Lambda_\rmn{cool}}
  = \frac{f_s X_\CR}{\mu_t}\,\frac{\vel_A kT}{n_e \Lambda_\rmn{cool}},
\end{equation}
where $\mu_t = n_t/n = (3X_\rmn{H}+1)/(5X_\rmn{H}+3)\simeq 0.48$ assuming
primordial element composition.

\begin{figure}
\begin{center}
\includegraphics[width=\columnwidth]{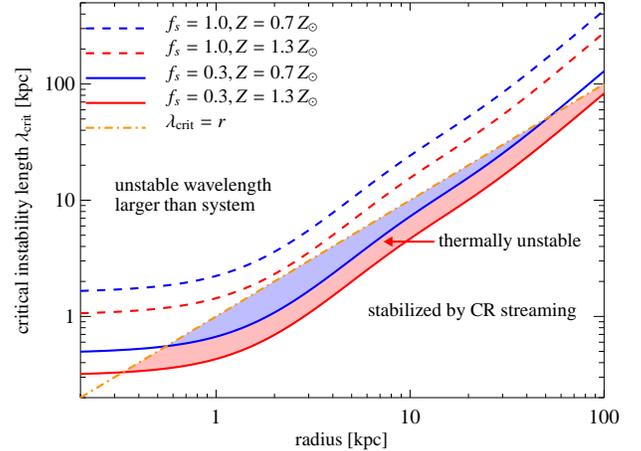}
\end{center}
\caption{Radial dependence of the critical instability length scale
  $\lambda_\rmn{crit}$ that is obtained by balancing the CR heating rate
  $\mathcal{H}_\CR$ with the radiative cooling rate $\mathcal{C}_\rmn{rad}$.  On
  scales $\lambda<\lambda_\rmn{crit}$, CR heating dominates over cooling and
  wipes out any temperature inhomogeneities. Gas parcels on scales
  $\lambda>\lambda_\rmn{crit}$ can become thermally unstable unless
  $\lambda_\rmn{crit}>r$, for which the unstable wavelengths cannot be supported
  by the system.  To compute the radiative cooling, we vary the abundance of the
  gas between $Z=0.7\,Z_\odot$ (blue) and $Z=1.3\,Z_\odot$ (red). We show one
  case without azimuthal fluctuations in the CR heating rate (dashed) and one
  with magnetic suppression by a factor $f_s=0.3$ (solid).}
\label{fig:lambda_crit}
\end{figure}

In Figure~\ref{fig:lambda_crit} we show $\lambda_\rmn{crit}$ as a function of
radius for the parameters of M87 used throughout, i.e., $\vel_A =
130~\rmn{km~s}^{-1}$, $X_\CR=0.31$, $\Lambda_\rmn{cool}(T_0, Z_0)$ of
Section~\ref{sec:global}, and the observed electron temperature and density
profiles of Equations \eqref{eq:ne} and \eqref{eq:Te}. The CR heating rate
without azimuthal fluctuations ($f_s=1$) yields a critical wavelength
$\lambda_\rmn{crit}>r$ at all radii, which implies that the unstable wavelengths
cannot be supported by the cluster core.

Since CRs stream anisotropically along magnetic fields, this can suppress the
heating flux into a given volume element depending on the magnetic connectivity
to the surroundings. In the case of an isotropically tangled small-scale
magnetic field, CR heating along a given direction is suppressed by a factor
$f_s\approx0.3$.  As a result, the minimum size of an unstable region at
10 kpc is $\lambda_\rmn{crit}=5$--8~kpc, depending on metallicity. However, such
a large region is unlikely to be exclusively heated from one direction, which
makes it challenging for cooling multi-phase gas to form unless a given region
is sufficiently magnetically insulated with a suppression factor $f_s\ll
1$. Since we have demonstrated in Figure~\ref{fig:heat} that CR heating globally
balances cooling, a magnetic suppression of heating into one region implies an
overheating in another. Because the gas is thermally unstable for temperatures in
the range of $(1$--$3)\times10^7$~K, the additional energy would be radiated away on
the radiative cooling time.

\subsection{The Emerging Physical Model of a ``Cooling Flow''}

The previous considerations enable us to construct a comprehensive ``cooling
flow'' model for M87, which may serve as a blueprint for other CC clusters. We
distinguish three different cases, each of which may be realized by a different
magnetic suppression factor. However, quantifying their relative abundance
requires appropriate numerical simulations.

{\em Case A.} In the absence of strong magnetic suppression of the CR heat flux,
a gas parcel is thermally stabilized by CR heating because the system cannot
support the unstable wavelengths, which are larger than the cluster radius.

{\em Case B.} For strong magnetic suppression of the CR heat flux, gas at
temperatures $T = (1$--$3)\times 10^7$~K is thermally unstable, starts to cool
radiatively, becomes denser than the ambient gas, and begins to sink toward the
center. Since CRs are largely confined to stay on individual field lines, which
are flux frozen into the contracting gas, the CR population is adiabatically
compressed. This implies an increasing CR heating rate $\mathcal{H}_\CR\propto
\rho^{\gamma+1/3}$ (Equation~\eqref{eq:HCR_OoM}), which however increases at a
slower rate than the cooling rate for $T=(1$--$3)\times 10^7$~K (see
Section~\ref{sec:local_stability}). The CRs transfer energy to the gas on a
timescale comparable to the Alfv{\'e}n crossing time over the CR pressure
gradient length, $\tau_A = \delta l/\vel_A \sim 75$~Myr for $\delta l = 10$~kpc.

If the gas parcel remains magnetically insulated, it cannot be supplied with a
fresh influx of CRs, which causes the CR heating to lag the radiative cooling,
implying eventually the formation of multi-phase gas and potentially stars,
provided that favorable conditions for their formation are met. Such a magnetic
suppression could be realized through processes that dynamically form coherently
aligned magnetic sheaths, which are wrapped around the gas parcel as a result of
either the action of magnetic draping \citep{2006MNRAS.373...73L,
  2007MNRAS.378..662R, 2008ApJ...677..993D}, the heat-flux-driven buoyancy
instability \citep{2008ApJ...673..758Q, 2008ApJ...677L...9P}, or through
radial collapse, which amplifies the azimuthal components of the magnetic field,
in combination with coherent motions of the cooling gas parcel so that the
azimuthal magnetic field becomes causally unconnected to its surroundings.

The comparably large critical wavelength of the instability,
$\lambda_\rmn{crit}$, at large radii (even for substantial magnetic suppression)
may make it difficult for thermal instability to condense out large gas clouds
(which requires a stable magnetic insulation on the correspondingly long
timescale $\tau_A$). Instead, it is more likely that cooling (multi-phase)
clouds form in the central regions at radii $\aplt3$~kpc because of the small
$\lambda_\rmn{crit}$. For this case, the presence of small pockets of
multi-phase gas at larger radii will then necessarily imply that they would have
to be uplifted by eddies driven by the rising radio lobes.

{\em Case C.} For moderate magnetic suppression of a given gas parcel at
$T=(1$--$3)\times 10^7$~K, CR heating will be able to approximately match
radiative cooling. That means we can consider the gas parcel to be in global
thermal equilibrium but subject to local thermal instability. If this global
thermal equilibrium can be maintained during the cooling of a necessarily large
region of wavelength $\lambda\aplt r$, and if the perturbations are in pressure
equilibrium with the surroundings, then the results of our stability analysis
apply. In particular, the gas collapse is halted at around $10^7$~K by the first
``island of stability'' situated at $T_\rmn{floor} = (0.7$--$1)\times
10^7$~K. Hence for {\em case C}, we predict a temperature ``floor'' at the
cluster center with typical values distributed within $T_\rmn{floor}$. The
insensitivity of the shape of the first ``island of stability'' on the various
parameters entering the instability criterion suggests that we found a universal
mechanism that sets a lower temperature floor in CC clusters (as long as the CR
pressure is sufficient to provide global thermal stability).

This prediction of a generic lower temperature floor at $T_\rmn{floor}$ is in
agreement with high-resolution X-ray observations of Virgo
\citep{2002A&A...386...77M, 2010MNRAS.407.2046M, 2010MNRAS.407.2063W},
suggesting a viable solution for the puzzling absence of a large amount of lower
temperature gas in Virgo that is expected for an unimpeded cooling flow. Despite
earlier claims of a universal temperature profile in CC clusters
\citep[e.g.,][]{2001MNRAS.328L..37A}, a temperature floor at a fixed fraction of
one-third of the virial radius \citep{2003ApJ...590..207P} appears not to be
supported by the latest data. Instead, using a large complete sample of 64
clusters with available high-quality X-ray data, \citet{2010A&A...513A..37H}
find a broad distribution of the central temperature drop of $T_0 / T_\rmn{vir}
= 0.2$--0.7 in CC clusters (see Figure 4 of \citealt{2010A&A...513A..37H}). Note
that $T_0$ is the temperature estimate of the central annulus and thus an upper
limit to the true central temperature since $T_0$ depends on angular resolution
and sensitivity of the cluster observation. The lower temperature in their
modified cooling flow model is in the range $(0.5$--$2)\times10^7$~K and
clusters around $10^7$~K. It is generally smaller than $T_0$, which is expected
if there are multi-temperature components present (see Figure 16 of
\citealt{2010A&A...513A..37H}). These findings are in perfect agreement with our
presented model, which implies a lower temperature floor around
$(0.7$--$1)\times10^7$~K as an attractor solution but certainly can stray from
this equilibrium value and assume higher central temperature values for an
intermittent period of overheating. A temperature floor similar to that observed
in M87 of around $5\times10^6$~K \citep{2010MNRAS.407.2063W} is found in
comparable high-resolution observations of the Perseus cluster, Centaurus
cluster, 2A 0335+096, A262, A3581, and HCG 62 \citep{2007MNRAS.381.1381S,
  2008MNRAS.385.1186S, 2009MNRAS.393...71S, 2010MNRAS.402..127S}. Clearly, the
lack of a larger amount of gas that is radiatively cooling below $5\times10^6$~K
implies the presence of heating.\footnote{Alternatively this may be explained by
  absorbing cold gas along the line of sight \citep{2013ApJ...767..153W}, which
  however would have to exhibit a fine-tuned spatial distribution so that its
  covering fraction exceeds unity without entering a cooling catastrophe.}

However, metastable gas in the high-density tail of density fluctuations can
cool below $T \simeq 7\times 10^6$~K and hence leave this ``island of
stability''. These dense fluctuations again become subject to thermal
instability, sourcing some of the observable multi-phase gas.  The collapse
could again be halted at the second and third ``island of stability'' just above
$10^6$~K and $10^5$~K. However, since cooling clouds at these temperatures are
associated with higher densities and smaller length scales, the amount of CR
heating now depends critically on the magnetic connectivity of the cloud with
the surroundings since that determines whether there will be a sufficient influx
of CRs responsible for sustaining the heating rate. These theoretical
considerations compare favorably to the observational result of a
multi-wavelength study of the emission-line nebulae located southeast of the
nucleus of M87, which reveals multi-phase material spanning a temperature range
of at least five orders of magnitude, from $\sim 100$~K to $\sim10^7$~K
\citep{2013ApJ...767..153W}. This material has a small total gas mass and has
most likely been uplifted by the AGN from the center of M87.

We emphasize that the success of our model came about without any tunable
parameters and only used input from non-thermal radio and gamma-ray
observations. The model implies that gas in the core region of Virgo is in
global equilibrium but susceptible to local thermal instability. Hence, this
provides a physical foundation of the phenomenological prescription used in
numerical simulations that study the interplay of thermal instability and
feedback regulation \citep{2012MNRAS.419.3319M, 2012MNRAS.420.3174S}.
Apparently, thermal instability can produce spatially extended multi-phase
filaments only when the instantaneous ratio of the thermal instability and
free-fall timescales falls below a critical threshold of about 10
\citep{2012MNRAS.420.3174S,2012ApJ...746...94G}.  There are also notable
differences since these prescriptions did not have the modulating effect of our
``islands of stability''. We may speculate that our model may then inherit some
of their successes including quantitative consistency with data on multi-phase
gas and star formation rates but holds the promise to cure the problem of the
absence of a (physical) temperature floor in these numerical experiments.

\section{Observational Predictions}
\label{sec:predictions}

This model makes several predictions for gamma-ray, radio, and SZ observations,
which we will discuss in turn. Some of the proposed observations may first be
possible in M87, owing to its immediate vicinity. To address the universality of
this heating mechanism, these observations are certainly worthwhile to be
pursued in other nearby CC clusters that show strong signs of AGN feedback.

\subsection{Gamma Rays}
\label{sec:gamma}

If the gamma-ray emission of the low state of M87 results from hadronic CRp--p
interactions then (1) the emission should be extended with a characteristic
angular profile worked out by \citet{2003A&A...407L..73P} for their isobaric CR
model, (2) the gamma-ray emission should not show any sign of time variability,
and (3) the emission should be characterized by the distinct spectral signature
of decaying pions around 70~MeV (in the differential spectrum). The triad of
these observations will provide strong support for the existence of CRps mixed
into the ICM and will allow us to measure their pressure contribution. Spatially
resolving the gamma-ray emission helps in estimating the large-scale CR pressure
gradient and its contribution to the CR-heating rate. However, this does not
shed light on the contribution from small-scale pressure fluctuations. The
comparably coarse angular resolution of \Fermi (68\% containment angle of
$12\arcmin$ above 10~GeV) precludes such a measurement in M87. In contrast, the
superior resolution of current and future Cerenkov telescopes (which is about a
few arc minutes with H.E.S.S.{\ }II above TeV energies and expected to be five
times better with CTA, approaching theoretical limits of $0\farcm3$ at 1 TeV;
see \citealt{2009arXiv0912.3742W}; \citealt{2006astro.ph..3076H}) is a
promising route toward measuring the pressure gradient of the high-energy CRps,
which, however, carry only a small amount of the total pressure that is dominated
by the GeV-energy regime.

We note that a critical prediction of this model is the similarity of the
spectral indices of the gamma-ray emission above GeV energies (which corresponds
to that of the CR population) and the injection index of the CRe population, as
probed by low-frequency radio observations of the core regions of Virgo
A. Similar observations of nearby AGNs in CC clusters are necessary to show the
universality of the presented mechanism.

\subsection{Radio Emission}
\label{sec:radio}

\begin{figure}
\begin{center}
\includegraphics[width=\columnwidth]{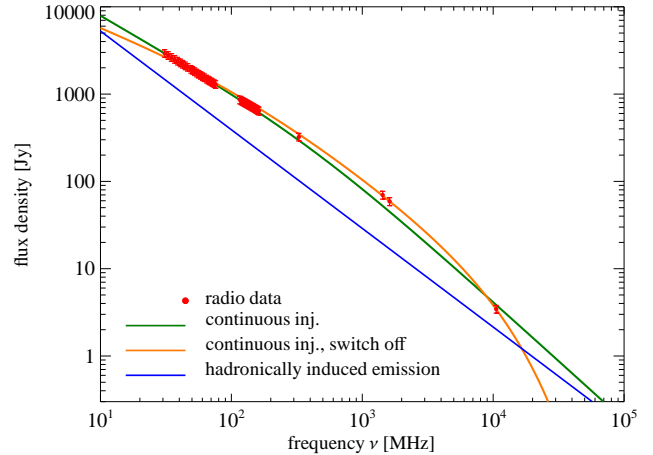}
\end{center}
\caption{Spectrum of the Virgo radio halo. We show the data for the halo without
  the central cocoon (red) as presented in \citet{2012A&A...547A..56D}. We
  compare the best-fit continuous injection model that allows for inverse
  Compton and synchrotron cooling (green, assuming $\alpha_\nu=0.86$) to a
  continuous injection model where the source has been switched off after a time
  interval (orange). In the latter model, the spectral index has been
  constrained to the injection index $\alpha_\nu=0.6$ that was measured for the
  central radio cocoon. Note that the expected contribution from the secondary
  radio emission that results from hadronic interactions of CRps with the
  ambient gas (blue) either dominates the external halo or makes up a large flux
  fraction at frequencies $\nu_\rmn{trans}\apgt20$~GHz.}
\label{fig:spectrum}
\end{figure}

The hadronic CRp--p reaction also implies the injection of secondary CRes that
contribute to the radio synchrotron emission in M87. Assuming steady state of
the hadronic injection of secondary CRes and synchrotron cooling, we obtain a
spectral index for the secondary radio emission of $\alpha_\nu=\alpha/2=1.13$
(adopting the CRp index $\alpha=2.26$ inferred in
Section~\ref{sec:Fermi}). Using the formalism of \citet{2008MNRAS.385.1211P}, we
calculate the hadronically induced radio halo emission within a spherical region
of radius 34~kpc (adopting the same parameters as in the rest of the paper,
namely, $X_\CR=0.31$, $\alpha=2.26$, $q=0.58$, and the model for the magnetic
field of Section~\ref{sec:CR}). At 1.4~GHz (10.55~GHz), this emission component
is expected to have a total flux density of 21.1~Jy (2.15~Jy), which falls short
of the observed halo emission by a factor of 6.6 (10).

However, the spatial profile of the expected hadronically induced emission is
very different from the observed radio halo emission, which is centrally more
concentrated. If we exclude a cylinder centered on the central radio cocoon of
radius 2.5~kpc (corresponding to $0\farcm51$), we obtain a flux density at
1.4~GHz (10.55~GHz) for the external hadronic halo of 19.8~Jy (2.0~Jy), which
still falls short of the observed emission of 69.8~Jy (3.4~Jy) that subtends the
same solid angle (see Figure~\ref{fig:spectrum}) but makes up a larger flux
fraction in comparison to that of the entire halo region. In particular, the
hadronically induced emission is expected to make up $\sim60\%$ of the observed
(external) halo emission at 10.55~GHz.

In Figure~\ref{fig:spectrum} we also show the continuous injection model
(green), which assumes an uninterrupted supply of fresh CRes from the central
source \citep{1970ranp.book.....P}. The CRes are subject to IC and synchrotron
cooling, which introduces a steepening of the power-law spectrum by $\Delta
\alpha_\nu = 0.5$ above a break frequency. The pitch angles of CRes are
conjectured to be continuously isotropized on a timescale shorter than the
radiative timescale \citep{1973A&A....26..423J}. Additionally shown is a similar
model, in which the source is switched off after a certain time
\citep[orange;][]{1994A&A...285...27K}. This causes the spectrum to drop
exponentially above a second break frequency, which corresponds to the cooling
time since the switch-off (see Appendix~\ref{sec:spectrum} for
details). Depending on the model of primary electrons, our model predicts that
the hadronically induced radio emission (blue) starts either to dominate the
halo emission or to make up a large fraction of the total emission at
frequencies $\nu_\rmn{trans}\apgt20$~GHz. This would reveal itself as a radio
spectral flattening at this transition frequency $\nu_\rmn{trans}$.  Clearly,
the possibility of such an emission component should be accounted for when
drawing conclusions about the CRe pitch-angle distribution in spectral aging
models of synchrotron and IC emitting CRes, especially if the high-frequency
data are sparse so that individual data points impart high statistical weights
in the fits.

If CRes and protons have a very different distribution within the radio halo
(which could be achieved through enhanced radiative losses of CRes in regions of
strong magnetic fields), there may be regions in which hadronically induced
emission prevails over the primary emission already at frequencies $\nu<20$~GHz.
Those regions should then also show a radio spectral flattening in comparison to
the adjacent primary emission-dominated regions.

\subsection{Faraday Rotation Measurements}
\label{sec:FRM}

Faraday rotation measurements of background sources or polarized regions of
Virgo A (central region, halo, or rising lobes) will provide a reliable estimate
not only of the spatial distribution of field strengths, but also about the
magnetic morphology. This is important for predicting the heating rates (that
scale with the Alfv{\'e}n velocity) and the microphysics of CR Alfv{\'e}n wave
heating, which depends on in situ alignments of magnetic fields with the CR
pressure gradients. In particular, CR release timescales from the radio lobes
may depend on the pre-existing morphology of the magnetic fields that
dynamically drape around a super-Alfv{\'e}nically rising lobe
\citep{2006MNRAS.373...73L, 2007MNRAS.378..662R}. This magnetic draping effect
stabilizes the interfaces against the shear that creates it in the first place
\citep{2007ApJ...670..221D}. However, magnetic draping can only occur if the
ratio of the integral length scale of the magnetic power spectrum $\lambda_B$
and the curvature radius $R$ of the radio lobe at the stagnation line obeys the
relation \citep[see Supplementary Information of][]{2010NatPh...6..520P}
\begin{equation}
  \frac{\lambda_B}{R} \apgt \M_A^{-1} = 
  \left(\M \sqrt{\frac{\gamma_\rmn{th}\beta}{2}}\right)^{-1}
  \simeq 0.25\, \M^{-1}\,\left(\frac{\beta}{20}\right)^{-1/2},
\end{equation}
where $\M$ and $\M_A$ are the sonic and Alfv{\'e}nic Mach numbers of the rising
lobes, $\gamma_\rmn{th}=5/3$, and $\beta$ is the plasma beta factor. The
violation of this relation could cause Kelvin--Helmholtz and Rayleigh--Taylor
instabilities to disrupt the lobes and to release CRs into the ICM. Higher order
Faraday rotation measure statistics are sensitive to the statistics of the
projected magnetic field distribution. Complementary to this are magnetic
draping studies at spiral galaxies within this central region
\citep{2010NatPh...6..520P} that provide a means of measuring the projected in
situ geometry of the magnetic field at the galaxies' position. Additionally,
those could be of paramount importance to provide indications for a spatially
preferred direction of magnetic fields that could hint at the presence of a
magnetic buoyancy instability.

\subsection{Sunyaev--Zel'dovich Observations}
\label{sec:SZ}

Cosmological cluster simulations of CR feedback by AGNs suggest a substantial
modification of the X-ray emissivity and the thermal SZ effect, with
modifications of the integrated Compton-$y$ parameter within the virial radius
up to 10\% \citep{2008MNRAS.387.1403S, 2013MNRAS.428.2366V}. High-resolution SZ
observations can even reveal the composition of the plasma within radio lobes
\citep{2005A&A...430..799P}.  Different dynamically dominating components within
radio lobes, which can generally range from relativistic to hot thermal
compositions, produce different SZ contrasts at the position of these lobes. A
relativistic filling implies an SZ cavity at the lobe positions. This is because
CRps do not Compton upscatter cosmic microwave background photons while CR
leptons that are in pressure equilibrium with the ambient ICM scatter only a
tiny fraction of background photons to X-ray energies in comparison to a
situation where the lobes would be filled with a slightly hotter than average
thermal population. In fact, the latter case would make the lobes almost
indistinguishable from the ambient cluster SZ effect. Both signatures are also
different from the hypothetical alternative of a hot but sub-relativistic
particle population. This would cause an SZ effect that is spectrally
distinguishable from the ordinary thermal SZ effect since it crosses through a
null above the canonical 217 GHz (but within the microwave
band). High-resolution ``SZ spectroscopy'' would be ideally suited to detect the
difference between these cases \citep{2005A&A...430..799P}.  However, the large
dynamic range required for such a detection in Virgo A will make such a
measurement very challenging.

\section{Discussion and Conclusions} 
\label{sec:C}

What is {\em the} dominant physical mechanism that quenches cooling flows in CC
clusters? Over the past decade, many different processes have been proposed. We
will discuss the pros and cons of the major contenders that are representative
of generic classes of heating models. For the first time, we have proposed a
comprehensive CC heating model that not only convincingly matches radio, X-ray,
and gamma-ray data for M87 but is also virtually free of tunable parameters. In
particular, our global and local stability analyses of the CR heating model
reveal generic aspects that may apply universally to CC clusters, provided that
the CR release timescales from the radio lobes are similar to that in M87.
\newline

\subsection{Overview of Solutions to the ``Cooling Flow Problem''}

Thermal conduction of heat from the hot outer cluster regions to the cooling
cores has been proposed to solve the problem \citep{2001ApJ...562L.129N,
  2003ApJ...582..162Z}. However, there are a number of drawbacks to this
proposal. (1) The strong temperature dependence of thermal conductivity
($\kappa\propto T^{5/2}$) makes this a very inefficient process for groups and
small clusters. (2) Thermal conduction ceases to be important if gas cools to
low temperatures and, as such, requires in some clusters a conductivity above
the classical Spitzer value, as well as a high degree of fine tuning
\citep{2004MNRAS.347.1130V, 2008ApJ...681..151C}. (3) The heat-flux-driven
buoyancy instability \citep{2008ApJ...673..758Q} causes the magnetic field lines
to align with the gravitational equipotential surfaces of the cluster in the
nonlinear stage of the instability \citep{2008ApJ...677L...9P}. At first sight,
this should suppress the inward transport of heat by more than an order of
magnitude and should reinforce the cooling flow problem. The instability,
however, is easily quenched by the presence of external turbulence at the level
of $\sim1\%$ of the thermal pressure \citep{2010ApJ...712L.194P,
  2010ApJ...713.1332R}, leaving the question of how much thermal conduction is
suppressed by the presence of magnetic fields in realistic cosmological
situations. Overall, thermal conduction could certainly aid in mitigating the
fast onset of cooling at intermediate cluster radii \citep{2011ApJ...740...28V}
but is unlikely to play a major role in the cluster centers. In particular, we
show in Appendix~\ref{sec:cond} that CR heating dominates over thermal
conduction for the entire CC region of the Virgo cluster.

Over the past years, AGN feedback has become the lead actor because of its
promising overall energetics and the observational confirmation that the nuclear
activity appears to be tightly coupled to a self-regulated feedback loop
\citep{2004ApJ...607..800B}. Guided by high-resolution {\em Chandra} X-ray
observations, which emphasize sharp density features in high-pass filtered maps
as a result of the gas being pushed around by AGN jets and lobes, many papers
have focused on the dissipation of $PdV$ work done by the inflating lobe on the
surroundings. The detailed heating mechanisms include sound waves, weak shocks,
and turbulence \citep[see][and references therein]{2012NJPh...14e5023M}. Since
sound waves and weak shocks are excited in the center by the AGN, their
outward-directed momentum flux in combination with the small dissipation
efficiency implies that only a small fraction of this energy will be used for
heating the central parts \citep[e.g.,][]{2012MNRAS.427.1482G}. Those parts,
however, have the lowest cooling time. Additionally, the implied heating rate is
very intermittent in space and time with an unknown duty cycle. Nevertheless,
these processes will certainly happen and (if the outburst energy is large
enough) may locally even dominate during some times
\citep{2005ApJ...628..629N}. As we argue in this paper, these processes are,
however, unlikely to be the most important heating source because they do not
provide a generic solution to the lower temperature floor observed in
CCs\footnote{Note that a temperature floor may also develop in the ``cold
  feedback mechanism'' provided that there exist nonlinear dense perturbations
  with a cooling time that is shorter than the AGN duty cycle. In this case
  those dense blobs may be able to cool to very low temperatures and feed the
  supermassive black hole or are expelled back to the ICM and heated by shocks
  and mixing \citep{2005ApJ...632..821P}.}  and they are not favored on
energetic grounds due to their small heating efficiency.

Assuming a relativistic filling of the lobes (for which we provide arguments in
Section~\ref{sec:content}), three-quarters of the total lobe enthalpy is stored in
internal energy after the end of the momentum-dominated jet phase. These light
relativistic lobes rise buoyantly and get subsequently disrupted by
Rayleigh--Taylor or Kelvin--Helmholtz instabilities. Once released, those CRps
stream at about the Alfv{\'e}n velocity with respect to the plasma rest frame
and excite Alfv\'en waves through the ``streaming instability''
\citep{1969ApJ...156..445K}. Damping of these waves transfers CR energy and
momentum to the thermal gas at a rate that scales with the product of the
Alfv{\'e}n velocity and the CR pressure gradient
\citep{2008MNRAS.384..251G}. While very plausible, there was the open question
of whether the CRp population has been mixed into the central ICM and whether it
is abundant enough to provide sufficiently strong heating to offset the cooling
catastrophe. In this paper, we answer both questions affirmatively for the case
of M87 on the basis of a detailed analysis of non-thermal observations.
Moreover, in addressing the issue of local thermal stability in this model, we
put forward an attractive solution to the problem of an apparent temperature
floor in CC clusters.

In particular, we observe in M87 that the lobes get disrupted at about one scale
height so that the density drops from the radius where the lobes start to rise
buoyantly to the current position at the radio halo boundary by a factor of
about $f = \rho_1/\rho_2 \simeq 10$.  Hence, the internal (relativistic) energy
is adiabatically cooled from $U_1 = 3 P_\rmn{th}V$ to $U_2 = U_1 f^{1-\gamma}
\simeq 1.5 P_\rmn{th}V$ at the time of release. A fraction of CRs gets caught in
the advective downdrafts that are driven by the buoyantly rising relativistic
lobes. This re-energizes the CRs adiabatically at the expense of mechanical
(e.g., turbulent) energy, thereby partially replenishing the energy initially
lost to the surroundings and making it available for CR heating at the dense ICM
regions with short cooling times. We end up with a fraction of $(1.5$--$3)\,
P_\rmn{th}V$ in CR and $(1$--$2.5)\, P_\rmn{th}V$ in mechanical energy. Since
streaming is an adiabatic process for the CR population \citep{Kulsrud}, a
fraction of the CR population can pass through several cycles of outward
streaming and downward advection, suggesting the physical realization of a larger
CR energy fraction closer to $3\, P_\rmn{th}V$ and a correspondingly smaller
mechanical energy fraction. The central cooling region delineated by the radio
halo is characterized by a sharp radio boundary, which suggests magnetic
confinement of CRs that can be realized by an azimuthal field
configuration. Hence, most of the CR energy will get dissipated within the halo
boundary, whereas it is not clear which fraction of the mechanical energy gets
dissipated in the central region. Among other uncertainties, this depends on the
partitioning of mechanical energy into turbulence and shocks.

\subsection{Cosmic-ray Heating as the Solution?}

The puzzling absence of fossil CRes implied by the LOFAR observations of M87 can
be naturally explained by the observed disruption of radio lobes during their
rise in the gravitational potential, e.g., through the action of
Kelvin--Helmholtz and Rayleigh--Taylor instabilities. This releases and mixes
CRes and protons into the ambient dense ICM. We argue that mixing is essential
in this picture so that the CRes have direct contact with the dense ambient
plasma, which causes CRes to efficiently cool through Coulomb interactions. This
implies a maximum CRe cooling timescale of $\simeq40$~Myr, which is similar to
the age of the Virgo A radio halo. A direct consequence of this interpretation
is that CRps are similarly mixed with the dense gas and must collide
inelastically with gas protons. This produces pions that decay into a steady
gamma-ray signature, which resembles that of the flux in the ``low state'' of
M87 observed by \Fermi and H.E.S.S. In particular, the CRp population (witnessed
through the gamma-ray emission) shares the same spectral index as the injected
CRe population in the core region of Virgo A, which we consider as strong
support for a common origin of the radio and gamma-ray emission.\footnote{As
  \citet{2012A&A...547A..56D}, we interpret the steeper CRe spectra in the halo
  of Virgo A as a result of a superposition of CRe spectra with different
  radiative cooling breaks that they attained during their transport from the
  core region, which is an effect that should leave the CRp spectra invariant.}
Hence, the hadronically induced gamma-ray emission enables us to estimate the CR
pressure, which has a CR-to-thermal pressure contribution of $X_\CR = 0.31$
(allowing for a conservative Coulomb cooling timescale of order the radio halo
age that modifies the low-energy part of the CRp distribution).

We show that streaming CRs heat the surrounding thermal plasma at a rate that
balances that of radiative cooling {\em on average} at each radius. Hence, the
thermal pressure profile is a consequence of the interplay of CR heating and
radiative cooling and should match that of the streaming CRs (hence justifying
in retrospect our assumption of a constant CR-to-thermal pressure in our
analysis). We argue that this similar radial scaling of the heating and cooling
rate profiles is not a coincidence in M87 but rather generically expected in
models of CR Alfv{\'e}n wave heating.

However, the {\em global} thermal equilibrium is {\em locally} unstable if the
gas temperature drops below $kT\simeq 3$~keV (for solar metallicity). The fate
of a gas parcel depends on its magnetic connectivity to the surroundings, which
may obey a yet unknown distribution function. Nevertheless, we can distinguish
three different cases.
\begin{enumerate}
\topsep0pt \itemsep2pt \parskip0pt \parsep0pt
\item For magnetically well-connected patches without appreciable fluctuations
  in the CR heating rate, temperature inhomogeneities are smoothed out on scales
  smaller than the cluster-centric radius. As a result, the system is stabilized
  because it cannot support the unstable wavelengths.

\item Contrarily, for strong magnetic suppression of the CR heat flux, the
  thermally unstable gas starts to cool radiatively, becomes denser than the
  ambient gas, and sinks toward the center. After an initial phase when the
  magnetically trapped CRs within the parcel deposit their energy to the gas, CR
  heating lags cooling. This eventually causes the formation of multi-phase gas
  and potentially stars.  At large radii, the critical wavelength
  $\lambda_\rmn{crit}$ above which thermal instability can operate is comparably
  large, thereby making it more likely for multi-phase gas to form further
  inward where $\lambda_\rmn{crit}$ is smaller. Thus, small clumps of
  multi-phase gas at larger radii would have to be uplifted by eddies driven by
  the rising radio lobes.

\item For the perhaps most probable case of moderate magnetic suppression, CR
  heating will be able to approximately match radiative cooling.  If this global
  thermal equilibrium can be maintained during the Lagrangian evolution of the
  cooling gas parcel that is in pressure equilibrium with its surroundings, then
  the gas collapse is halted at around 1~keV by an ``island of stability.''
  There, the CR heating rate has a steeper dependence on temperature than the
  radiative cooling rate.  Increasing (decreasing) the temperature of a
  perturbed parcel of gas implies a stronger cooling (heating) that brings the
  parcel back to the background temperature.  This is consistent with the
  observed temperature floor of 0.7--1~keV seen in the X-ray observations of the
  center of Virgo. Since this result is insensitive to significant variations of
  the CR pressure support, this may suggest a viable solution to the ``cooling
  flow problem'' in other CC clusters (as long as the CR heating rate is
  sufficient to provide global thermal stability). However, the high-density
  tail of density fluctuations that is associated with temperatures
  $kT\aplt0.7$~keV becomes again subject to thermal instability, thereby
  providing a plausible mechanism for the formation of some of the observed
  cooling multi-phase medium in and around M87.
\end{enumerate}
We note that the success of the CR heating model came about without any tunable
parameters and {\em only} used input from radio and gamma-ray observations. We
present several testable predictions of this model that include gamma-ray,
radio, and SZ observations.

We emphasize the importance of non-thermal observations to establish this
physical picture of AGN feedback and to pinpoint the details of this heating
mechanism in M87. While tempting, we caution against directly extrapolating our
result to other CC clusters. To faithfully do so, we would need similar
non-thermal observations to establish that CRs are injected and mixed into the
ambient ICM and to estimate the CR pressure contribution that determines the CR
heating rate.  Since the residency time of the CRps in the central core region
is longer than the age of the radio halo, CR heating can take place even without
(high-)frequency radio halo emission.

In particular, CR release timescales from the radio lobes are crucial and may
depend on the pre-existing morphology of the magnetic fields that dynamically
drape around the rising lobes \citep{2007MNRAS.378..662R}. Those draped fields
provide stability in the direction of the magnetic field against the shear that
creates it if the magnetic coherence scale is large enough for draping to be
established in the first place. Future Faraday rotation measure studies and
galaxy draping observations \citep{2010NatPh...6..520P} may help to clarify this
point.

We hope that this work motivates fruitful observational and theoretical efforts
toward consolidating the presented picture and to establish non-thermal
observations of the underlying high-energy processes as a driver for the
understanding of the evolution of cluster cores and perhaps the modulation of
star formation in elliptical galaxies.

\acknowledgments I thank Ewald Puchwein and Volker Springel for critical reading
of the manuscript and helpful suggestions. I acknowledge useful suggestions and
comments from Eugene Churazov, Heino Falcke, Paul Nulsen, Peng Oh, Frank Rieger,
Ellen Zweibel, and an anonymous referee that helped to improve the paper.  I
thank Francesco de Gasperin for the permission to show the LOFAR image and
spectral data of M87 and Matteo Murgia for explaining details of the best-fit
radio emission models. I gratefully acknowledge financial support of the Klaus
Tschira Foundation.

\begin{appendix}

\section{Thermal Stability in the Presence of a Relativistic Fluid}
\label{sec:stability}

\subsection{Balancing Bremsstrahlung and Line Cooling by Generic Heating Mechanisms}
\label{sec:general}

Consider a fluid subject to external heating and cooling. The generalized Field
criterion \citep{1965ApJ...142..531F, 1986ApJ...303L..79B} for thermal
instability in the presence of isobaric perturbations is given by
\begin{equation}
  \label{eq:Field1}
  \left.\frac{\partial (\calL / T)}{\partial T}\right|_P < 0.
\end{equation}
Here $\calL$ is the net loss function in units of erg s$^{-1}$ g$^{-1}$ defined
such that $\rho\calL=\mathcal{C}-\mathcal{H}$, where $\mathcal{C}$ and
$\mathcal{H}$ are the cooling and heating rates per unit volume. $\calL$ is
generally a function of gas mass density $\rho$, temperature $T$, and possibly
space and time. Earlier local stability analyses neglected CR pressure support
\citep{1991ApJ...377..392L} or were restricted to a cooling function of the
interstellar medium at temperatures around $10^4$~K \citep{2013ApJ...767...87W}.
We study the instability in the presence of a relativistic fluid and generalize
the derivation by \citet{2008ApJ...681..151C} to also include cooling by line
emission in addition to thermal bremsstrahlung, which will turn out to be
important for putting the complete picture in place (in particular for cooling
flows in clusters).

A gas in thermal equilibrium ($\rho\calL=0$) obeys the following identity,
\begin{equation}
  \label{eq:Field2}
  \left.\frac{\partial (\calL / T)}{\partial T}\right|_P =
  \frac{1}{\rho T}\left[
    \left.\frac{\partial (\rho\calL)}{\partial T}\right|_\rho
    + \left.\frac{\partial \rho}{\partial T}\right|_P
    \left.\frac{\partial (\rho\calL)}{\partial \rho}\right|_T
  \right].
\end{equation}
The first and third derivatives on the right-hand side are governed by the
heating and cooling processes under consideration, while the second depends on
the various sources of pressure support. We consider pressure support provided
by thermal gas and a relativistic fluid (dominated by CRps):
\begin{equation}
  \label{eq:P}
P = P_\rmn{th} + P_\CR = K_1 \rho T + P_\CR(\rho),
\end{equation}
where $K_1$ is a constant and $P_\CR$ is only a function of density. We use the
following identity
\begin{equation}
  \label{eq:identity1}
  \left.\frac{\partial \rho}{\partial T}\right|_P = 
  -\left.\frac{\partial P}{\partial T}\right|_\rho 
  \left( \left.\frac{\partial P}{\partial \rho}\right|_T \right)^{-1}
\end{equation}
to obtain
\begin{equation}
  \label{eq:identity2}
  \left.\frac{\partial \rho}{\partial T}\right|_P = 
  -\frac{K_1\rho}{K_1 T + \left.\left(\partial P_\CR/\partial\rho\right)\right|_T}
  =-\frac{\rho}{T}\frac{P_\rmn{th}}{P_\rmn{th} + \rho\left.\left(\partial P_\CR/\partial\rho\right)\right|_T}
  =-\frac{\rho}{T}\frac{1}{1 + X_\CR\left.\left(\partial \ln P_\CR/\partial\ln\rho\right)\right|_T},
\end{equation}
where we defined $X_\CR \equiv P_\CR/P_\rmn{th}$. 

We now assume that the heating has a power-law dependence on $\rho$ and $T$ and
account for cooling by thermal bremsstrahlung and line emission to get
\begin{equation}
  \label{eq:loss function}
  \rho\calL = \mathcal{C}_\rmn{rad} - \mathcal{H}
  = \rho^2 \left[\Lambda_b (T,Z) + \Lambda_l(T,Z)\right]
  - \Gamma_0 \rho^\beta T^{-1/2+\delta},
\end{equation}
where $\Lambda_b(T,Z) = \Lambda_0 Z^2 T^{1/2}$ and $\Lambda_l(T,Z)$ are the
cooling functions due to bremsstrahlung and line emission, respectively. Both
depend on metallicity and temperature. (For brevity, we omit the arguments in the
following.) $\Lambda_0$ and $\Gamma_0$ are constants, and we define the
identical power-law indices $\delta$ and $\beta$ for the heating as in
\citet{2008ApJ...681..151C} to ease comparison with their
results. Using the relation for thermal equilibrium ($\rho\calL=0$), we
have
\begin{equation}
  \label{eq:deriv1}
  \left.\frac{\partial (\rho\calL)}{\partial T}\right|_\rho = 
  \Lambda_b\rho^2T^{-1}
  \left[
      1-\delta + \frac{\Lambda_l}{\Lambda_b}
      \left(
        \left.\frac{\partial\ln \Lambda_l}{\partial \ln T}\right|_\rho
        + \frac{1}{2} - \delta
      \right)
    \right]
\end{equation}
and
\begin{equation}
  \label{eq:deriv2}
  \left.\frac{\partial (\rho\calL)}{\partial \rho}\right|_T =
  \Lambda_b\rho\,
      (2-\beta) \left(1 + \frac{\Lambda_l}{\Lambda_b} \right).
\end{equation}
Combining those with Equations~\eqref{eq:Field2} and \eqref{eq:identity2}, we finally obtain
\begin{equation}
  \label{eq:deriv3}
  \left.\frac{\partial (\rho\calL)}{\partial T}\right|_P =
  \Lambda_b\rho T^{-2}
  \left[
      1-\delta + \frac{\Lambda_l}{\Lambda_b}
      \left(
        \left.\frac{\partial\ln \Lambda_l}{\partial \ln T}\right|_\rho
        + \frac{1}{2} - \delta
      \right)
      - \frac{2-\beta}{1+X_\CR\gamma}\,
      \left(1 + \frac{\Lambda_l}{\Lambda_b} \right)
    \right].
\end{equation}
To simplify the instability criterion, we define the logarithmic density slope
of the CR pressure $\gamma$, the logarithmic temperature slope of the line
cooling function $\chi$, and the ratio of line-to-bremsstrahlung cooling,
$\eta$:
\begin{equation}
  \label{eq:abbrev}
  \gamma \equiv \left.\frac{\partial\ln P_\CR}{\partial \ln \rho}\right|_T,\quad\quad
  \chi = \chi(T,Z) \equiv \left.\frac{\partial\ln \Lambda_l}{\partial \ln T}\right|_\rho,\quad\,\mbox{and}\quad
  \eta = \eta(T,Z) \equiv \frac{\Lambda_l}{\Lambda_b}.
\end{equation}
Thus, the gas is thermally unstable if the discriminant $\mathcal{D}$ is
negative:
\begin{equation}
  \label{eq:instability1}
  \mathcal{D} \equiv 1-\delta + \eta
  \left(
    \chi + \frac{1}{2} - \delta
  \right)
  - \frac{(2-\beta) (1 + \eta )}{1+X_\CR\gamma} < 0.
\end{equation}
For $\eta=0$ we recover the result of \citet{2008ApJ...681..151C} if we identify
$X_\CR\equiv1/\alpha'$ in their notation. Clearly, the presence of line cooling
($\eta>0$) can substantially modify the instability criterion, depending on
the temperature slope $\chi$.

\subsection{Thermal Stability of a Radiatively Cooling Gas Heated by
  Cosmic-ray-excited Alfv{\'e}n Waves}
\label{sec:CRA}

Now, we specify this instability criterion to our case at hand of CR Alfv{\'e}n
wave heating. We assume the existence of small-scale tangled magnetic fields
that result, e.g., from magnetohydrodynamic turbulence. Then, magnetic flux
freezing during the contraction of unstable cooling gas ensures adiabatic
compression of CRs.  On the Alfv{\'e}nic crossing time that is of relevance
for our stability analysis, the CR population does not suffer non-adiabatic
losses such as hadronic interactions \citep[e.g.,][]{2007A&A...473...41E}. In
particular, CR streaming is an adiabatic process for the CR population
\citep{Kulsrud}, implying locally $P_\CR\propto\rho^\gamma$ and yielding
\begin{equation}
\label{eq:HCR}
\mathcal{H}_\CR
  = - \bvel_A\bcdot\boldsymbol{\nabla} P_\CR = \Gamma_0 \rho^\beta T^{-1/2+\delta} 
  \propto \rho^{\gamma+1/3}.
\end{equation}
Here we identified the characteristic size of our unstable clump with an
exponential pressure scale height $h$ so that the local gradient length scales
as $\delta l \sim h \propto\rho^{-1/3}$ and we assume $\vel_A=\rmn{const}$. In
practice, $\vel_A$ may inherit a weak density scaling depending on the topology
of the magnetic field that is compressed or expanded. We explore the impact of a
density-dependent $\vel_A$ on the global and local thermal equilibrium in
Appendix~\ref{sec:v_A} but defer dynamical simulations of this interacting
CR-magnetohydrodynamic system to future work.  Equation~\eqref{eq:HCR} defines
the power-law indices $\beta=\gamma+1/3$ and $\delta=1/2$ for the heating
rate. We emphasize that this local adiabatic behavior of $P_\CR$ does not hold
for the equilibrium distribution that is subject to gain and loss processes,
which modify the ``CR entropy'' (i.e., its probability density of the
microstates). In fact, we argued in the main part of the paper that CR
Alfv{\'e}n wave heating establishes an equilibrium distribution that is
characterized by $X_\CR = P_\CR/P_\rmn{th}\simeq \rmn{const.}$ {\em if} the
dominant heating source is provided by CRs.

To estimate the CR adiabatic exponent, we remind ourselves of the definition of
the CR pressure of a CR power-law population defined in Equation~\eqref{eq:fCR},
which is given by
\begin{equation}
\label{eq:Pcr}
P_\CR = 
\frac{m_p c^2}{3}\,\int_0^\infty  d p\,\frac{d^4N}{dp\,d^3x}
\,\frac{\vel}{c}\,p =
\frac{C\,m_p c^2}{6} \, 
  \B_{\frac{1}{1+q^2}} \left( \frac{\alpha-2}{2},\frac{3-\alpha}{2} \right),
\end{equation}
where $\vel/c = p/\sqrt{1+p^2}$ is the dimensionless velocity of the CR particle
and $\B_x(a,b)$ denotes the incomplete beta function.  The local adiabatic
exponent of the CR population is obtained by taking the logarithmic derivative
of the CR pressure at constant CR entropy $S_\CR$
\begin{equation}
\label{eq:gammaCR}
\gamma \equiv \left.\frac{\partial \log P_\CR}{\partial \log \rho}\right|_{S_\CR}=
\frac{\rho}{P_\CR}
\left(
  \frac{\partial P_\CR}{\partial C}\frac{\partial C}{\partial \rho}
  + \frac{\partial P_\CR}{\partial q}\frac{\partial q}{\partial \rho}
\right) 
= \frac{\alpha + 2}{3} - \frac{2}{3}\,
q^{2-\alpha}\, \frac{\vel(q)}{c}\,
\left[
  \B_{\frac{1}{1+q^2}} 
 \left( \frac{\alpha-2}{2},\frac{3-\alpha}{2}\right) 
\right]^{-1}\simeq1.37
\end{equation}
where we used typical values for M87 of $\alpha=2.26$ and $q=0.58$ as derived
in Section~\ref{sec:Fermi}.

The ratio of line-to-bremsstrahlung cooling at 1~keV for a gas with solar
abundance is approximately $\eta\simeq3.6$ \citep[see Figure~8
of][]{1993ApJS...88..253S}, assuming collisional ionization equilibrium. We can
now rewrite Equation~\eqref{eq:instability1} to get a constraint for the
temperature slope of the cooling function $\chi$ for which the gas can become
thermally unstable,
\begin{equation}
  \label{eq:instability2}
  \chi < -\frac{1}{2\eta}
  + \frac{\left(2-\beta\right)(1+\eta)}{\left(1+X_\CR\gamma\right)\eta}\simeq 0.13,
  \mbox{ for 1 keV and }Z=Z_\odot.
\end{equation}
Hence, if the cooling function is sufficiently flat or even inverted (as it can
be the case in the line cooling regime), we have thermal instability. In the
bremsstrahlung regime, on the contrary, the gas is in thermal equilibrium, since
the criterion for instability
\begin{equation}
  \label{eq:instability3}
  \delta > \frac{X_\CR\gamma-1+\beta}{1+X_\CR\gamma}\simeq0.8
\end{equation}
cannot be fulfilled for CR Alfv{\'e}n wave heating that does not depend on
temperature ($\delta=1/2$).

\section{Varying Alfv{\'e}n speed}
\label{sec:v_A}

\subsection{Impact on Global Thermal Equilibrium}

\begin{figure}
\begin{center}
\includegraphics[width=0.495\columnwidth]{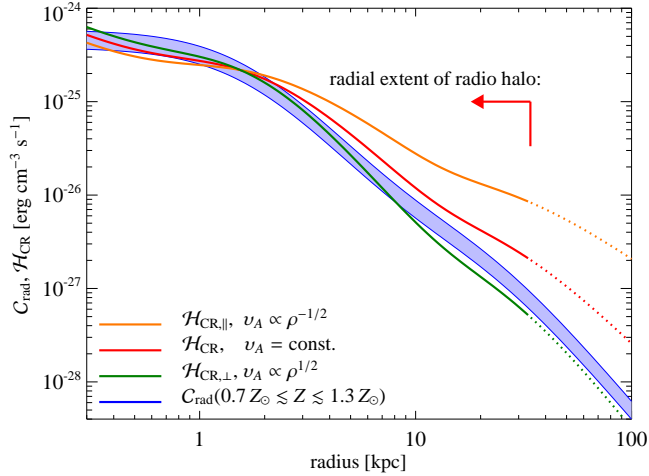}
\end{center}
\caption{Impact of a varying $\vel_A$ on global thermal equilibrium.  We compare
  the radiative cooling rate (blue) to the CR Alfv{\'e}n wave heating rate
  ($\mathcal{H}_\rmn{CR}$) due to the averaged pressure profile and pressure
  fluctuations (assuming weak shocks of typical Mach number $\M=1.1$). We
  contrast the standard case of $\vel_A=\rmn{const.}$ (red) to the case of a
  collapse along $\mathbfit{B}$ (implying $\vel_A \propto \rho^{-1/2}$, orange)
  and perpendicular to $\mathbfit{B}$ (implying $\vel_A \propto \rho^{1/2}$,
  green). }
\label{fig:v_A}
\end{figure}

So far, we assumed that the Alfv{\'e}n speed is independent of radius (or
density). Whether this is true depends in practice on the magnetic topology of a
collapsing (or expanding) fluid parcel. Parameterizing $B\propto
\rho^{\alpha_B}$, we have $\alpha_B=0$ for collapse along $\mathbfit{B}$
(implying $\vel_{A,\parallel} \propto \rho^{-1/2}$) and $\alpha_B=1$ for
collapse perpendicular to $\mathbfit{B}$ (implying $\vel_{A,\perp} \propto
\rho^{1/2}$). By construction, we have at the center $c_s/\vel_A\simeq
4$. Hence, at the halo boundary of Virgo A, we obtain for the ratio of
sound-to-Alfv{\'e}n speed $c_s/\vel_{A,\parallel}
=\sqrt{\beta\gamma_\rmn{th}/2}\simeq 1.5$ and $c_s/\vel_{A,\perp}\simeq 26$
(where $\beta$ is the plasma beta factor and $\gamma_\rmn{th}=5/3$). This
neglects a small-scale dynamo that could modify the $B-\rho$ scaling
furthermore. Hence, our discussed case of $\vel_A=\rmn{const.}$ is the geometric
mean of these extreme cases and implies that field components perpendicular and
parallel to the direction of collapse are present. We normalize the CR heating
at a radius of $r_0 \simeq 1.5~\rmn{kpc}$ that corresponds to the radial extent
of the central radio cocoon, which provides rather homogeneous conditions for
field estimates based on the minimum energy conditions. Note that for
simplicity, we hold the pressure gradient fixed while varying the Alfv{\'e}n
speed with radius, implicitly assuming a sufficiently large magnetic
connectivity in radial direction. Hence, the CR heating rate at
$r_\rmn{max}=35~\rmn{kpc}$ could be at most overestimated by a factor of
$\vel_A/\vel_{A,0} = (\rho/\rho_0)^{-1/2}\simeq 3.8$, which is about the
enhancement factor of the CR heating rate with $\vel_A=\rmn{const.}$ over the
cooling rate at the halo boundary for $Z = 0.7~Z_\odot$ (see
Figure~\ref{fig:v_A}).  We can interpret this result twofold. Either we have
$\mathcal{H}_\CR> \mathcal{C}_\rmn{rad}$ and the halo boundary is currently
adiabatically expanding in the cluster atmosphere or we have global thermal
equilibrium, i.e., $\mathcal{H}_\CR= \mathcal{C}_\rmn{rad}$.

\subsection{Impact on Local Stability Analysis}

\begin{figure*}
\begin{center}
\includegraphics[width=0.495\columnwidth]{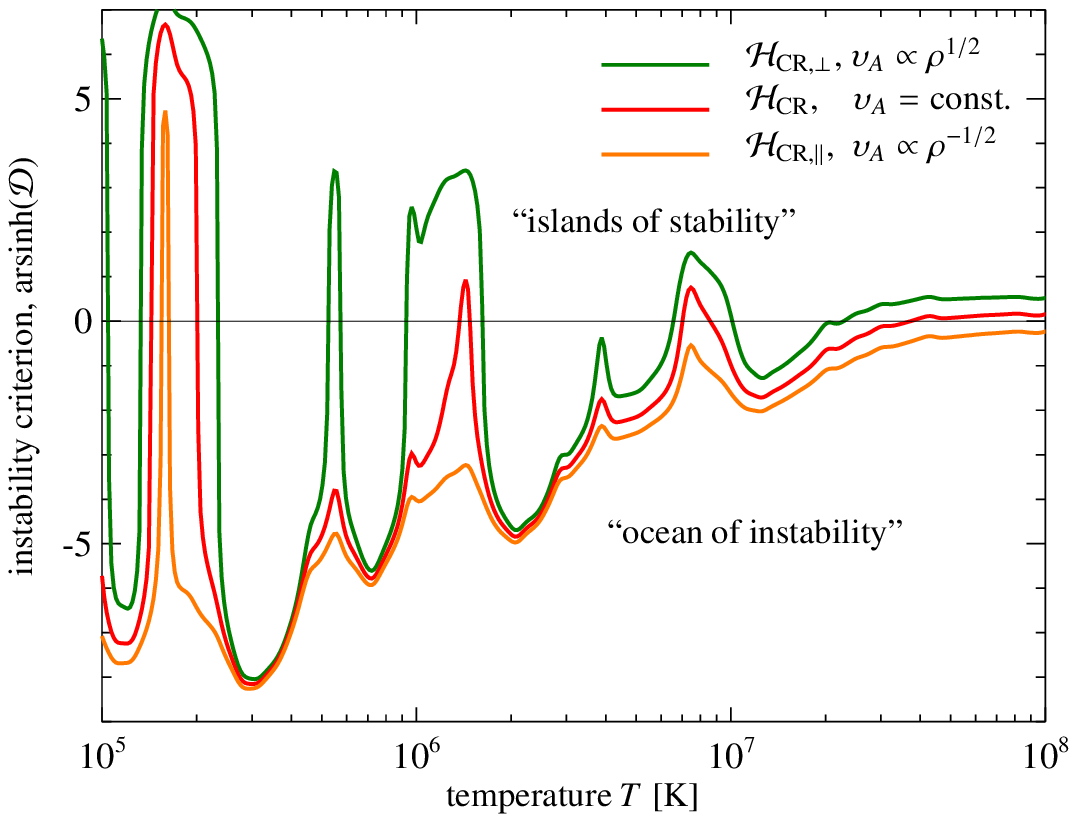}
\includegraphics[width=0.495\columnwidth]{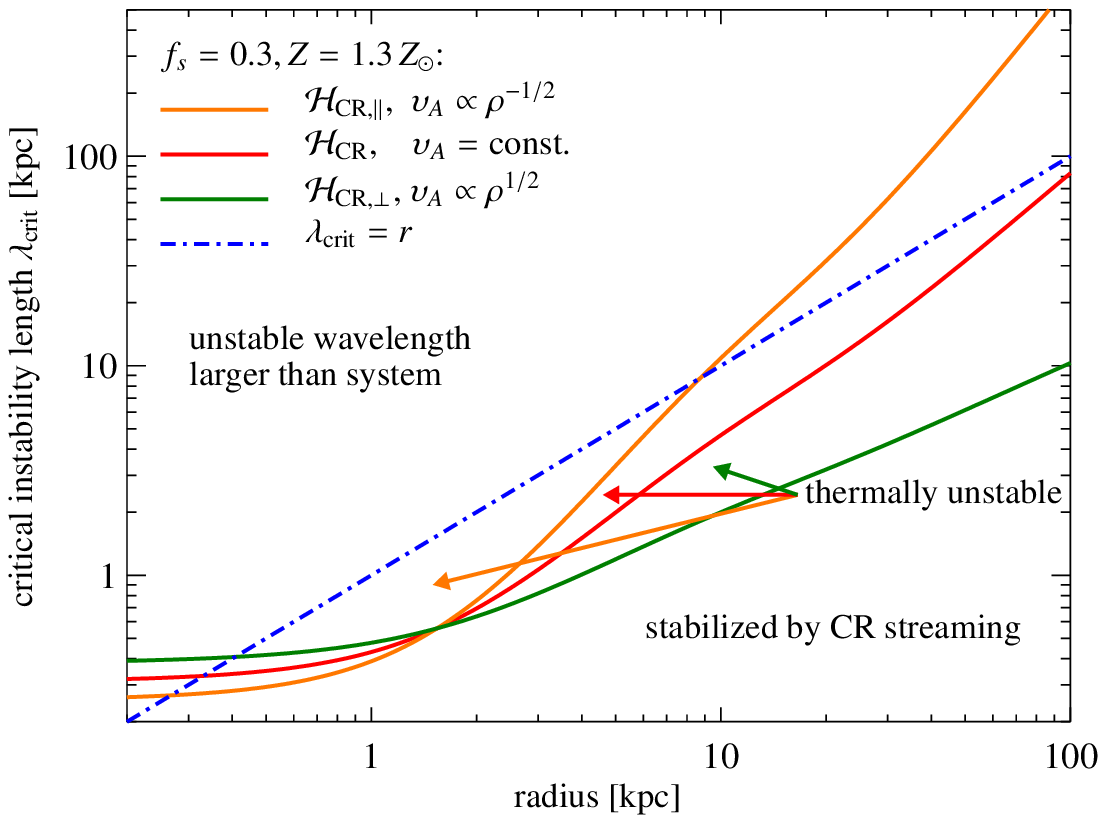}
\end{center}
\caption{Impact of a varying $\vel_A$ on the local stability analysis. Left: we
  show the local instability criterion, $\mathcal{D}\propto\left.\partial
    (\calL/T)/\partial T\right|_P$ as a function of temperature for parameters
  appropriate for M87. In comparison to the standard case of
  $\vel_A=\rmn{const.}$ (red), the ``islands of stability'' get more prominent
  for collapse perpendicular to $\mathbfit{B}$ (implying $\vel_A \propto
  \rho^{1/2}$, green). On the contrary, for gas collapsing along $\mathbfit{B}$
  (implying $\vel_A \propto \rho^{-1/2}$, orange), the global thermal
  equilibrium is always locally unstable for $T>2\times10^5$~K. Right: we show
  the radial dependence of the critical instability length scale
  $\lambda_\rmn{crit}$ that is obtained by balancing the CR heating rate
  ($\mathcal{H}_\CR$) with the radiative cooling rate ($\mathcal{C}_\rmn{rad}$)
  for $Z=1.3\,Z_\odot$ and $f_s=0.3$.  On scales $\lambda<\lambda_\rmn{crit}$,
  CR heating dominates over cooling and stabilizes the system. Gas parcels can
  become thermally unstable on scales $\lambda>\lambda_\rmn{crit}$ unless
  $\lambda_\rmn{crit}>r$, for which the unstable wavelengths cannot be supported
  by the system.  For a gas parcel of fixed metallicity and magnetic
  connectivity outside the center, $\lambda_\rmn{crit}$ decreases for collapse
  perpendicular to $\mathbfit{B}$ ($\vel_A \propto \rho^{1/2}$, green), i.e.,
  smaller patches can become unstable and start to cool until they are
  stabilized at $T=10^7$~K (see left panel). }
\label{fig:v_A2}
\end{figure*}

How does a varying Alfv{\'e}n speed impact on the local stability analysis?  We
reconsider Equation~\eqref{eq:HCR} and adopt the density dependencies of
$\vel_A$ discussed in the previous section, i.e., $\alpha_B = \{0, 0.5, 1\}$ to
obtain
\begin{equation}
\label{eq:HCR2}
\mathcal{H}_\CR
  = - \bvel_A\bcdot\boldsymbol{\nabla} P_\CR = \Gamma_0 \rho^\beta T^{-1/2+\delta} 
  \propto \rho^{\gamma+1/3\pm1/2}, 
  ~\rmn{i.e.,}\quad
  \beta-\gamma=\left\{-\frac{1}{6}, \frac{2}{6}, \frac{5}{6}\right\},
\end{equation}
where the first and last value of $\beta$ corresponds to collapse parallel and
perpendicular to $\mathbfit{B}$, respectively. Following the discussion
surrounding Equation~\eqref{eq:C_OoM}, we can understand the character of local
thermal stability by comparing the (modified) slope of the CR heating rate to
that of the radiative cooling rate. We find that the slope of the parallel CR
heating rate is always steeper than that of the radiative cooling rate for
$T>2\times10^5$~K, which implies local instability throughout this temperature
regime. Note that this case is characterized by $\mathcal{H}_\CR>
\mathcal{C}_\rmn{rad}$ (Figure~\ref{fig:v_A}) so that the instability analysis
formally does not apply and would require lowering $X_\CR$ until
$\mathcal{H}_\CR= \mathcal{C}_\rmn{rad}$.

In contrast, the ``islands of stability'' get more prominent for perpendicular
collapse in comparison to the standard case of $\vel_A=\rmn{const.}$ and even a
new ``island of stability'' emerges out of the ``ocean of instability'' at
around $5\times10^5$~K (left panel of Figure~\ref{fig:v_A2}). It is interesting to
note that the latter case either represents the quiescently saturating state of
the heat-flux-driven buoyancy instability \citep{2008ApJ...673..758Q,
  2008ApJ...677L...9P} or could be obtained through radial collapse, which
amplifies the azimuthal components of the magnetic field. Since these processes
are likely to generate or amplify a sufficiently large azimuthal field
component, the realization of the pure radial instability case has a small
probability.

In the right panel of Figure~\ref{fig:v_A2}, we address how a varying Alfv{\'e}n
speed changes the critical instability length scale ($\lambda_\rmn{crit}$) that
is obtained by balancing the CR heating rate with the radiative cooling rate. On
scales $\lambda<\lambda_\rmn{crit}$, CR heating dominates over cooling and wipes
out temperature inhomogeneities. Gas parcels can become thermally unstable on
scales $\lambda>\lambda_\rmn{crit}$ provided that the unstable wavelengths can
be supported by the system. For a gas parcel of fixed metallicity and magnetic
connectivity outside the center, $\lambda_\rmn{crit}$ decreases for collapse
perpendicular to $\mathbfit{B}$ ($\vel_A \propto \rho^{1/2}$), i.e., smaller
regions can become unstable and start to cool until they are stabilized at
$T=10^7$~K (see left panel of Figure~\ref{fig:v_A2}). In the opposite case of
collapsing gas along $\mathbfit{B}$ ($\vel_A \propto \rho^{-1/2}$) we have
formally local instability for $T>2\times10^5$~K. However, for gas of
metallicity $Z=1.3\,Z_\odot$ and a magnetic suppression factor $f_s=0.3$ of the
heating rate, cooling is nevertheless stabilized for $r\gtrsim10$~kpc since the
unstable wavelengths cannot be supported by the system.

\section{Thermal Conduction versus Cosmic-ray Heating}
\label{sec:cond}

\begin{figure}
\begin{center}
\includegraphics[width=0.495\columnwidth]{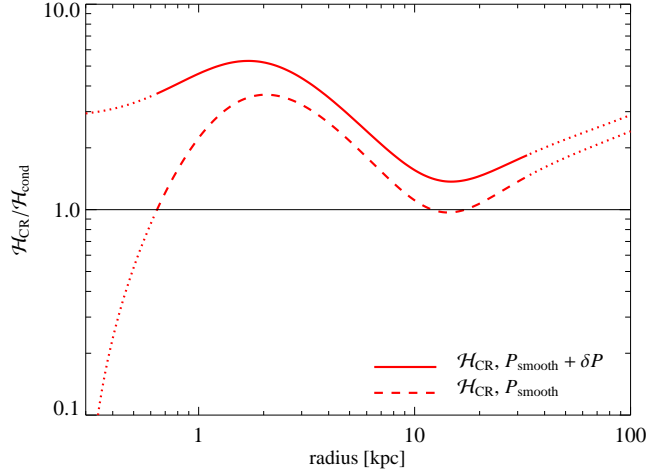}
\end{center}
\caption{We show the ratio of the heating rates due to CR streaming
    ($\mathcal{H}_\CR$) and thermal conduction ($\mathcal{H}_\rmn{cond}$) as a
    function of radius for parameters appropriate for the CC of
    Virgo/M87. We compare $\mathcal{H}_\CR$ due to the azimuthally averaged
    pressure profile that ``smoothes'' out the CR pressure gradient (dashed red)
    and the $\mathcal{H}_\CR$ profile, where we additionally account for
    pressure fluctuations due to weak shocks of typical Mach number $\M=1.1$
    (solid red). CR heating dominates over thermal conduction throughout the
    Virgo CC region. The CR heating rates become very uncertain outside
    the boundary of the radio halo of Virgo A while the temperature and pressure
    profiles inside of 0.6 kpc (corresponding to $0\farcm1$) are not well
    constrained by the data (as indicated with dotted red lines in both
    cases).}
\label{fig:cond}
\end{figure}

Electron thermal conduction transports energy into a fluid element at a rate
\begin{equation}
\label{eq:cond}
\hat{\mathcal{H}}_\rmn{cond} = r^2 \kappa(T)\,\mathbfit{n}\bcdot\boldsymbol{\nabla} T, 
  ~\rmn{where} \quad \kappa(T) = 6\times 10^{-7} T^{5/2} f_s
  ~\rmn{erg}~\s^{-1}~\K^{-1}~\cm^{-1}
\end{equation}
is the Spitzer conduction coefficient, $\mathbfit{n}$ is the normal of the
surface area element, and $f_s$ is a magnetic suppression factor that is
determined by the magnetic connectivity of the fluid parcel with the hotter
surroundings. Similarly, streaming CRs deposit energy into the same fluid
element at a rate $\hat{\mathcal{H}}_\CR = r^3 f_s \mathcal{H}_\CR$, where
$\mathcal{H}_\CR$ is given by Equation~\eqref{eq:CR}. Taking the ratio of the
heating rates due to CR streaming and thermal conduction enables us to assess
the relative importance of both mechanisms. First, we note that this ratio is
independent of $f_s$ provided that the magnetic connectivity of a fluid element
to the external hotter regions is on average equal to the magnetic connectivity
to the external regions with a higher CR pressure. In Figure~\ref{fig:cond}, we
show the ratio $\mathcal{H}_\CR /\mathcal{H}_\rmn{cond}$ as a function of
cluster-centric radius for parameters appropriate for M87. We observe a decline
of this ratio for the radial range of $2~\rmn{kpc}<r< 10$~kpc due to strong
temperature dependence of thermal conduction, which makes this mechanism
relatively more important in comparison to CR heating (although conduction
always stays subdominant). For radii $r\gtrsim10$~kpc, we observe the opposite
behavior, i.e., CR heating becomes again more important than thermal conduction
because of the steepening of the thermal pressure gradient in comparison to the
diminishing temperature gradient at these scales. Especially for the case that
additionally accounts for pressure fluctuations in the CR heating rate,
conduction can at best aid in mitigating the onset of cooling at intermediate
cluster radii while CR heating clearly dominates over the entire scale of the CC
in Virgo/M87.

However, the strong temperature dependence of thermal conduction may increase
its heating rate over the CR heating rate in hotter clusters. Nevertheless,
thermal conduction is always locally unstable to temperature perturbations
because of this strong temperature dependence. To order of magnitude,
$\mathcal{H}_\rmn{cond} \propto \kappa(T) T / r^2\propto P^{2/3} T^{17/6}$,
which has a much steeper temperature dependence in comparison to the radiative
cooling function so that a locally unstable fluid element cannot be conductively
stabilized (for isobaric perturbations). Since thermal conduction is not
self-regulating, there is no mechanism to arrest catastrophic cooling in this
picture. Hence, even if conductive heating dominates CR heating at some high
temperature, the much shallower temperature dependence of CR heating inevitably
ensures that it dominates the heating budget of a cooling fluid element at lower
temperatures and may be responsible for setting the observed temperature floor.

\section{Radio Spectra of Cooling Electrons}
\label{sec:spectrum}

Here we present approximate functional forms for the radio spectra of cooling
CRes that we discuss in Section~\ref{sec:radio}. In particular, we compare the
continuous injection model (CI), which assumes an uninterrupted supply of fresh
CRes from the central source, to a similar model that also allows for the source
to be switched off after a certain time (CIoff). IC and synchrotron cooling of
CRes modifies their power-law injection spectrum, yielding a spectral index
above the cooling break that is steeper by $\Delta \alpha_e = 1$.  Hence, the
radio synchrotron spectrum steepens by $\Delta \alpha_\nu = 0.5$ above the break
frequency, which corresponds to the total age of the source. Switching the
source off after some time causes the spectrum to drop exponentially above a
second (higher) break frequency, which corresponds to the cooling time since
switch off:
\begin{eqnarray}
\label{eq:spectrum}
  j_\rmn{CI}(\nu) &=& S_\rmn{CI} \left(\frac{\nu}{\nu_0}\right)^{-\alpha_{\nu,\rmn{fit}}}\,
  \left(1 + \frac{\nu}{\nu_b}\right)^{-0.5}, \\
  j_\rmn{CIoff}(\nu) &=& S_\rmn{CIoff} \left(\frac{\nu}{\nu_0}\right)^{-\alpha_{\nu,\rmn{inj}}}\,
  \left(1 + \frac{\nu}{\nu_1}\right)^{-0.5}\, \exp\left(-\frac{\nu}{\nu_2}\right).
\end{eqnarray}
The parameters that fit the data of the radio halo without the central radio
cocoon of M87 are given by the best-fit low-frequency spectral index
$\alpha_{\nu,\rmn{fit}} = 0.86$ and the injection spectral index of the central
cocoon $\alpha_{\nu,\rmn{inj}}=0.6$. We normalize the spectra at $\nu_0 =
10$~MHz with $S_\rmn{CI} = 8\times 10^3$~Jy and $S_\rmn{CIoff} =
6\times10^3$~Jy. Finally, the various break frequencies are given by $(\nu_b,
\nu_1, \nu_2) = (0.4, 0.1, 11)$~GHz. We refrain from discussing the CI model
with the spectral index fixed to the injection spectral index of the central
cocoon because it does not match the data at 10.5~GHz
\citep{2012A&A...547A..56D}.
\newline

\end{appendix}

\bibliography{paper.bib}

\begin{thebibliography}{94}
\expandafter\ifx\csname natexlab\endcsname\relax\def\natexlab#1{#1}\fi

\bibitem[{{Abdo} {et~al.}(2009){Abdo}, {Ackermann}, {Ajello}, {Atwood},
  {Axelsson}, {Baldini}, {Ballet}, {Barbiellini}, {Bastieri}, {Bechtol},
  {Bellazzini}, {Berenji}, {Blandford}, {Bloom}, {Bonamente}, {Borgland},
  {Bregeon}, {Brez}, {Brigida}, {Bruel}, {Burnett}, {Caliandro}, {Cameron},
  {Cannon}, {Caraveo}, {Casandjian}, {Cavazzuti}, {Cecchi}, {{\c C}elik},
  {Charles}, {Cheung}, {Chiang}, {Ciprini}, {Claus}, {Cohen-Tanugi},
  {Colafrancesco}, {Conrad}, {Costamante}, {Cutini}, {Davis}, {Dermer}, {de
  Angelis}, {de Palma}, {Digel}, {Donato}, {Silva}, {Drell}, {Dubois},
  {Dumora}, {Edmonds}, {Farnier}, {Favuzzi}, {Fegan}, {Finke}, {Focke},
  {Fortin}, {Frailis}, {Fukazawa}, {Funk}, {Fusco}, {Gargano}, {Gasparrini},
  {Gehrels}, {Georganopoulos}, {Germani}, {Giebels}, {Giglietto}, {Giommi},
  {Giordano}, {Giroletti}, {Glanzman}, {Godfrey}, {Grenier}, {Grondin},
  {Grove}, {Guillemot}, {Guiriec}, {Hanabata}, {Harding}, {Hayashida}, {Hays},
  {Horan}, {J{\'o}hannesson}, {Johnson}, {Johnson}, {Johnson}, {Johnson},
  {Kamae}, {Katagiri}, {Kataoka}, {Kawai}, {Kerr}, {Kn{\"o}dlseder}, {Kocian},
  {Kuss}, {Lande}, {Latronico}, {Lemoine-Goumard}, {Longo}, {Loparco}, {Lott},
  {Lovellette}, {Lubrano}, {Madejski}, {Makeev}, {Mazziotta}, {McConville},
  {McEnery}, {Meurer}, {Michelson}, {Mitthumsiri}, {Mizuno}, {Moiseev},
  {Monte}, {Monzani}, {Morselli}, {Moskalenko}, {Murgia}, {Nolan}, {Norris},
  {Nuss}, {Ohsugi}, {Omodei}, {Orlando}, {Ormes}, {Ozaki}, {Paneque},
  {Panetta}, {Parent}, {Pelassa}, {Pepe}, {Pesce-Rollins}, {Piron}, {Porter},
  {Rain{\`o}}, {Rando}, {Razzano}, {Reimer}, {Reimer}, {Reposeur}, {Ritz},
  {Rochester}, {Rodriguez}, {Romani}, {Roth}, {Ryde}, {Sadrozinski},
  {Sambruna}, {Sanchez}, {Sander}, {Saz Parkinson}, {Scargle}, {Sgr{\`o}},
  {Shaw}, {Smith}, {Smith}, {Spandre}, {Spinelli}, {Strickman}, {Suson},
  {Tajima}, {Takahashi}, {Tanaka}, {Taylor}, {Thayer}, {Thompson}, {Tibaldo},
  {Torres}, {Tosti}, {Tramacere}, {Uchiyama}, {Usher}, {Vasileiou}, {Vilchez},
  {Waite}, {Wang}, {Winer}, {Wood}, {Ylinen}, {Ziegler}, {Harris}, {Massaro},
  \& {Stawarz}}]{2009ApJ...707...55A}
{Abdo}, A.~A., {et~al.} 2009, \apj, 707, 55

\bibitem[{{Abramowski} {et~al.}(2012){Abramowski}, {Acero}, {Aharonian},
  {Akhperjanian}, {Anton}, {Balzer}, {Barnacka}, {Barres de Almeida},
  {Becherini}, {Becker}, \& et~al.}]{2012ApJ...746..151A}
{Abramowski}, A., {et~al.} 2012, \apj, 746, 151

\bibitem[{{Ackermann} {et~al.}(2010){Ackermann}, {Ajello}, {Allafort},
  {Baldini}, {Ballet}, {Barbiellini}, {Bastieri}, {Bechtol}, {Bellazzini},
  {Blandford}, {Blasi}, {Bloom}, {Bonamente}, {Borgland}, {Bouvier}, {Brandt},
  {Bregeon}, {Brigida}, {Bruel}, {Buehler}, {Buson}, {Caliandro}, {Cameron},
  {Caraveo}, {Carrigan}, {Casandjian}, {Cavazzuti}, {Cecchi}, {{\c C}elik},
  {Charles}, {Chekhtman}, {Cheung}, {Chiang}, {Ciprini}, {Claus},
  {Cohen-Tanugi}, {Colafrancesco}, {Cominsky}, {Conrad}, {Dermer}, {de Palma},
  {Silva}, {Drell}, {Dubois}, {Dumora}, {Edmonds}, {Farnier}, {Favuzzi},
  {Frailis}, {Fukazawa}, {Funk}, {Fusco}, {Gargano}, {Gasparrini}, {Gehrels},
  {Germani}, {Giglietto}, {Giordano}, {Giroletti}, {Glanzman}, {Godfrey},
  {Grenier}, {Grondin}, {Guiriec}, {Hadasch}, {Harding}, {Hayashida}, {Hays},
  {Horan}, {Hughes}, {Jeltema}, {J{\'o}hannesson}, {Johnson}, {Johnson},
  {Johnson}, {Kamae}, {Katagiri}, {Kataoka}, {Kerr}, {Kn{\"o}dlseder}, {Kuss},
  {Lande}, {Latronico}, {Lee}, {Lemoine-Goumard}, {Longo}, {Loparco}, {Lott},
  {Lovellette}, {Lubrano}, {Madejski}, {Makeev}, {Mazziotta}, {Michelson},
  {Mitthumsiri}, {Mizuno}, {Moiseev}, {Monte}, {Monzani}, {Morselli},
  {Moskalenko}, {Murgia}, {Naumann-Godo}, {Nolan}, {Norris}, {Nuss}, {Ohsugi},
  {Omodei}, {Orlando}, {Ormes}, {Ozaki}, {Paneque}, {Panetta}, {Pepe},
  {Pesce-Rollins}, {Petrosian}, {Pfrommer}, {Piron}, {Porter}, {Profumo},
  {Rain{\`o}}, {Rando}, {Razzano}, {Reimer}, {Reimer}, {Reposeur}, {Ripken},
  {Ritz}, {Rodriguez}, {Romani}, {Roth}, {Sadrozinski}, {Sander}, {Saz
  Parkinson}, {Scargle}, {Sgr{\`o}}, {Siskind}, {Smith}, {Spandre}, {Spinelli},
  {Starck}, {Stawarz}, {Strickman}, {Strong}, {Suson}, {Tajima}, {Takahashi},
  {Takahashi}, {Tanaka}, {Thayer}, {Thayer}, {Tibaldo}, {Tibolla}, {Torres},
  {Tosti}, {Tramacere}, {Uchiyama}, {Usher}, {Vandenbroucke}, {Vasileiou},
  {Vilchez}, {Vitale}, {Waite}, {Wang}, {Winer}, {Wood}, {Yang}, {Ylinen}, \&
  {Ziegler}}]{2010ApJ...717L..71A}
{Ackermann}, M., {et~al.} 2010, \apjl, 717, L71

\bibitem[{{Ackermann} {et~al.}(2013){Ackermann}, {Ajello}, {Albert},
  {Allafort}, {Atwood}, {Baldini}, {Ballet}, {Barbiellini}, {Bastieri},
  {Bechtol}, {Bellazzini}, {Bloom}, {Bonamente}, {Bottacini}, {Brandt},
  {Bregeon}, {Brigida}, {Bruel}, {Buehler}, {Buson}, {Caliandro}, {Cameron},
  {Caraveo}, {Cavazzuti}, {Chaves}, {Chiang}, {Chiaro}, {Ciprini}, {Claus},
  {Cohen-Tanugi}, {Conrad}, {D'Ammando}, {de Angelis}, {de Palma}, {Dermer},
  {Digel}, {Drell}, {Drlica-Wagner}, {Favuzzi}, {Franckowiak}, {Funk}, {Fusco},
  {Gargano}, {Gasparrini}, {Germani}, {Giglietto}, {Giordano}, {Giroletti},
  {Godfrey}, {Gomez-Vargas}, {Grenier}, {Guiriec}, {Gustafsson}, {Hadasch},
  {Hayashida}, {Hewitt}, {Hughes}, {Jeltema}, {J{\'o}hannesson}, {Johnson},
  {Kamae}, {Kataoka}, {Kn{\"o}dlseder}, {Kuss}, {Lande}, {Larsson},
  {Latronico}, {Llena Garde}, {Longo}, {Loparco}, {Lovellette}, {Lubrano},
  {Mayer}, {Mazziotta}, {McEnery}, {Michelson}, {Mitthumsiri}, {Mizuno},
  {Monzani}, {Morselli}, {Moskalenko}, {Murgia}, {Nemmen}, {Nuss}, {Ohsugi},
  {Orienti}, {Orlando}, {Ormes}, {Perkins}, {Pesce-Rollins}, {Piron}, {Pivato},
  {Rain{\`o}}, {Rando}, {Razzano}, {Razzaque}, {Reimer}, {Reimer}, {Ruan},
  {S{\'a}nchez-Conde}, {Schulz}, {Sgr{\`o}}, {Siskind}, {Spandre}, {Spinelli},
  {Storm}, {Strong}, {Suson}, {Takahashi}, {Thayer}, {Thayer}, {Thompson},
  {Tibaldo}, {Tinivella}, {Torres}, {Troja}, {Uchiyama}, {Usher},
  {Vandenbroucke}, {Vianello}, {Vitale}, {Winer}, {Wood}, {Zimmer}, {Pfrommer},
  \& {Pinzke}}]{2013arXiv1308.5654T}
---. 2013, arXiv:1308.5654

\bibitem[{{Aharonian} {et~al.}(2006){Aharonian}, {Akhperjanian}, {Bazer-Bachi},
  {Beilicke}, {Benbow}, {Berge}, {Bernl{\"o}hr}, {Boisson}, {Bolz}, {Borrel},
  {Braun}, {Brown}, {B{\"u}hler}, {B{\"u}sching}, {Carrigan}, {Chadwick},
  {Chounet}, {Coignet}, {Cornils}, {Costamante}, {Degrange}, {Dickinson},
  {Djannati-Ata{\"i}}, {Drury}, {Dubus}, {Egberts}, {Emmanoulopoulos},
  {Espigat}, {Feinstein}, {Ferrero}, {Fiasson}, {Fontaine}, {Funk}, {Funk},
  {F{\"u}{\ss}ling}, {Gallant}, {Giebels}, {Glicenstein}, {Goret},
  {Hadjichristidis}, {Hauser}, {Hauser}, {Heinzelmann}, {Henri}, {Hermann},
  {Hinton}, {Hoffmann}, {Hofmann}, {Holleran}, {Hoppe}, {Horns},
  {Jacholkowska}, {de Jager}, {Kendziorra}, {Kerschhaggl}, {Kh{\'e}lifi},
  {Komin}, {Konopelko}, {Kosack}, {Lamanna}, {Latham}, {Le Gallou},
  {Lemi{\`e}re}, {Lemoine-Goumard}, {Lenain}, {Lohse}, {Martin},
  {Martineau-Huynh}, {Marcowith}, {Masterson}, {Maurin}, {McComb}, {Moulin},
  {de Naurois}, {Nedbal}, {Nolan}, {Noutsos}, {Orford}, {Osborne}, {Ouchrif},
  {Panter}, {Pelletier}, {Pita}, {P{\"u}hlhofer}, {Punch}, {Ranchon},
  {Raubenheimer}, {Raue}, {Rayner}, {Reimer}, {Ripken}, {Rob}, {Rolland},
  {Rosier-Lees}, {Rowell}, {Sahakian}, {Santangelo}, {Saug{\'e}}, {Schlenker},
  {Schlickeiser}, {Schr{\"o}der}, {Schwanke}, {Schwarzburg}, {Schwemmer},
  {Shalchi}, {Sol}, {Spangler}, {Spanier}, {Steenkamp}, {Stegmann}, {Superina},
  {Tam}, {Tavernet}, {Terrier}, {Tluczykont}, {van Eldik}, {Vasileiadis},
  {Venter}, {Vialle}, {Vincent}, {V{\"o}lk}, {Wagner}, \&
  {Ward}}]{2006Sci...314.1424A}
{Aharonian}, F., {et~al.} 2006, Science, 314, 1424

\bibitem[{{Aleksi{\'c}} {et~al.}(2010){Aleksi{\'c}}, {Antonelli}, {Antoranz},
  {Backes}, {Baixeras}, {Balestra}, {Barrio}, {Bastieri}, {Becerra
  Gonz{\'a}lez}, {Bednarek}, {Berdyugin}, {Berger}, {Bernardini}, {Biland},
  {Bock}, {Bonnoli}, {Bordas}, {Borla Tridon}, {Bosch-Ramon}, {Bose}, {Braun},
  {Bretz}, {Britzger}, {Camara}, {Carmona}, {Carosi}, {Colin}, {Commichau},
  {Contreras}, {Cortina}, {Costado}, {Covino}, {Dazzi}, {De Angelis}, {De Cea
  del Pozo}, {De los Reyes}, {De Lotto}, {De Maria}, {De Sabata}, {Delgado
  Mendez}, {Doert}, {Dom{\'{\i}}nguez}, {Dominis Prester}, {Dorner}, {Doro},
  {Elsaesser}, {Errando}, {Ferenc}, {Fonseca}, {Font}, {Galante},
  {Garc{\'{\i}}a L{\'o}pez}, {Garczarczyk}, {Gaug}, {Godinovic}, {Hadasch},
  {Herrero}, {Hildebrand}, {H{\"o}hne-M{\"o}nch}, {Hose}, {Hrupec}, {Hsu},
  {Jogler}, {Klepser}, {Kr{\"a}henb{\"u}hl}, {Kranich}, {La Barbera}, {Laille},
  {Leonardo}, {Lindfors}, {Lombardi}, {Longo}, {L{\'o}pez}, {Lorenz},
  {Majumdar}, {Maneva}, {Mankuzhiyil}, {Mannheim}, {Maraschi}, {Mariotti},
  {Mart{\'{\i}}nez}, {Mazin}, {Meucci}, {Miranda}, {Mirzoyan}, {Miyamoto},
  {Mold{\'o}n}, {Moles}, {Moralejo}, {Nieto}, {Nilsson}, {Ninkovic}, {Orito},
  {Oya}, {Paiano}, {Paoletti}, {Paredes}, {Partini}, {Pasanen}, {Pascoli},
  {Pauss}, {Pegna}, {Perez-Torres}, {Persic}, {Peruzzo}, {Prada}, {Prandini},
  {Puchades}, {Puljak}, {Reichardt}, {Rhode}, {Rib{\'o}}, {Rico}, {Rissi},
  {R{\"u}gamer}, {Saggion}, {Saito}, {Salvati}, {S{\'a}nchez-Conde},
  {Satalecka}, {Scalzotto}, {Scapin}, {Schultz}, {Schweizer}, {Shayduk},
  {Shore}, {Sierpowska-Bartosik}, {Sillanp{\"a}{\"a}}, {Sitarek}, {Sobczynska},
  {Spanier}, {Spiro}, {Stamerra}, {Steinke}, {Struebig}, {Suric}, {Takalo},
  {Tavecchio}, {Temnikov}, {Terzic}, {Tescaro}, {Teshima}, {Torres}, {Vankov},
  {Wagner}, {Zabalza}, {Zandanel}, {Zanin}, {Zapatero}, {Pfrommer}, {Pinzke},
  {En{\ss}lin}, {Inoue}, {Ghisellini}, \& {MAGIC
  Collaboration}}]{2010ApJ...710..634A}
{Aleksi{\'c}}, J., {et~al.} 2010, \apj, 710, 634

\bibitem[{{Aleksi{\'c}} {et~al.}(2012){Aleksi{\'c}}, {Alvarez}, {Antonelli},
  {Antoranz}, {Asensio}, {Backes}, {Barres de Almeida}, {Barrio}, {Bastieri},
  {Becerra Gonz{\'a}lez}, {Bednarek}, {Berdyugin}, {Berger}, {Bernardini},
  {Biland}, {Blanch}, {Bock}, {Boller}, {Bonnoli}, {Borla Tridon}, {Braun},
  {Bretz}, {Ca{\~n}ellas}, {Carmona}, {Carosi}, {Colin}, {Colombo},
  {Contreras}, {Cortina}, {Cossio}, {Covino}, {Dazzi}, {de Angelis}, {de
  Caneva}, {de Cea Del Pozo}, {de Lotto}, {Delgado Mendez}, {Diago Ortega},
  {Doert}, {Dom{\'{\i}}nguez}, {Dominis Prester}, {Dorner}, {Doro},
  {Eisenacher}, {Elsaesser}, {Ferenc}, {Fonseca}, {Font}, {Fruck},
  {Garc{\'{\i}}a L{\'o}pez}, {Garczarczyk}, {Garrido}, {Giavitto},
  {Godinovi{\'c}}, {Gozzini}, {Hadasch}, {H{\"a}fner}, {Herrero}, {Hildebrand},
  {H{\"o}hne-M{\"o}nch}, {Hose}, {Hrupec}, {Jogler}, {Kellermann}, {Klepser},
  {Kr{\"a}henb{\"u}hl}, {Krause}, {Kushida}, {La Barbera}, {Lelas}, {Leonardo},
  {Lewandowska}, {Lindfors}, {Lombardi}, {L{\'o}pez}, {L{\'o}pez},
  {L{\'o}pez-Oramas}, {Lorenz}, {Makariev}, {Maneva}, {Mankuzhiyil},
  {Mannheim}, {Maraschi}, {Mariotti}, {Mart{\'{\i}}nez}, {Mazin}, {Meucci},
  {Miranda}, {Mirzoyan}, {Mold{\'o}n}, {Moralejo}, {Munar-Adrover},
  {Niedzwiecki}, {Nieto}, {Nilsson}, {Nowak}, {Orito}, {Paiano}, {Paneque},
  {Paoletti}, {Pardo}, {Paredes}, {Partini}, {Perez-Torres}, {Persic},
  {Peruzzo}, {Pilia}, {Pochon}, {Prada}, {Prada Moroni}, {Prandini}, {Puerto
  Gimenez}, {Puljak}, {Reichardt}, {Reinthal}, {Rhode}, {Rib{\'o}}, {Rico},
  {R{\"u}gamer}, {Saggion}, {Saito}, {Saito}, {Salvati}, {Satalecka},
  {Scalzotto}, {Scapin}, {Schultz}, {Schweizer}, {Shayduk}, {Shore},
  {Sillanp{\"a}{\"a}}, {Sitarek}, {Snidaric}, {Sobczynska}, {Spanier}, {Spiro},
  {Stamatescu}, {Stamerra}, {Steinke}, {Storz}, {Strah}, {Sun}, {Suri{\'c}},
  {Takalo}, {Takami}, {Tavecchio}, {Temnikov}, {Terzi{\'c}}, {Tescaro},
  {Teshima}, {Tibolla}, {Torres}, {Treves}, {Uellenbeck}, {Vankov}, {Vogler},
  {Wagner}, {Weitzel}, {Zabalza}, {Zandanel}, {Zanin}, {MAGIC Collaboration},
  {Pfrommer}, \& {Pinzke}}]{2012A&A...541A..99A}
---. 2012, \aap, 541, A99

\bibitem[{{Allen} {et~al.}(2001){Allen}, {Schmidt}, \&
  {Fabian}}]{2001MNRAS.328L..37A}
{Allen}, S.~W., {Schmidt}, R.~W., \& {Fabian}, A.~C. 2001, \mnras, 328, L37

\bibitem[{{Arlen} {et~al.}(2012){Arlen}, {Aune}, {Beilicke}, {Benbow},
  {Bouvier}, {Buckley}, {Bugaev}, {Byrum}, {Cannon}, {Cesarini}, {Ciupik},
  {Collins-Hughes}, {Connolly}, {Cui}, {Dickherber}, {Dumm}, {Falcone},
  {Federici}, {Feng}, {Finley}, {Finnegan}, {Fortson}, {Furniss}, {Galante},
  {Gall}, {Godambe}, {Griffin}, {Grube}, {Gyuk}, {Holder}, {Huan}, {Hughes},
  {Humensky}, {Imran}, {Kaaret}, {Karlsson}, {Kertzman}, {Khassen}, {Kieda},
  {Krawczynski}, {Krennrich}, {Lee}, {Madhavan}, {Maier}, {Majumdar},
  {McArthur}, {McCann}, {Moriarty}, {Mukherjee}, {Nelson}, {O'Faol{\'a}in de
  Bhr{\'o}ithe}, {Ong}, {Orr}, {Otte}, {Park}, {Perkins}, {Pohl}, {Prokoph},
  {Quinn}, {Ragan}, {Reyes}, {Reynolds}, {Roache}, {Ruppel}, {Saxon},
  {Schroedter}, {Sembroski}, {Skole}, {Smith}, {Telezhinsky}, {Te{\v s}i{\'c}},
  {Theiling}, {Thibadeau}, {Tsurusaki}, {Varlotta}, {Vivier}, {Wakely}, {Ward},
  {Weinstein}, {Welsing}, {Williams}, {Zitzer}, {Pfrommer}, \&
  {Pinzke}}]{2012ApJ...757..123A}
{Arlen}, T., {et~al.} 2012, \apj, 757, 123

\bibitem[{{Balbus}(1986)}]{1986ApJ...303L..79B}
{Balbus}, S.~A. 1986, \apjl, 303, L79

\bibitem[{{Battaglia} {et~al.}(2010){Battaglia}, {Bond}, {Pfrommer}, {Sievers},
  \& {Sijacki}}]{2010ApJ...725...91B}
{Battaglia}, N., {Bond}, J.~R., {Pfrommer}, C., {Sievers}, J.~L., \& {Sijacki},
  D. 2010, \apj, 725, 91

\bibitem[{{B{\^i}rzan} {et~al.}(2004){B{\^i}rzan}, {Rafferty}, {McNamara},
  {Wise}, \& {Nulsen}}]{2004ApJ...607..800B}
{B{\^i}rzan}, L., {Rafferty}, D.~A., {McNamara}, B.~R., {Wise}, M.~W., \&
  {Nulsen}, P.~E.~J. 2004, \apj, 607, 800

\bibitem[{{B{\^i}rzan} {et~al.}(2012){B{\^i}rzan}, {Rafferty}, {Nulsen},
  {McNamara}, {R{\"o}ttgering}, {Wise}, \& {Mittal}}]{2012MNRAS.427.3468B}
{B{\^i}rzan}, L., {Rafferty}, D.~A., {Nulsen}, P.~E.~J., {McNamara}, B.~R.,
  {R{\"o}ttgering}, H.~J.~A., {Wise}, M.~W., \& {Mittal}, R. 2012, \mnras, 427,
  3468

\bibitem[{{Blanton} {et~al.}(2003){Blanton}, {Sarazin}, \&
  {McNamara}}]{2003ApJ...585..227B}
{Blanton}, E.~L., {Sarazin}, C.~L., \& {McNamara}, B.~R. 2003, \apj, 585, 227

\bibitem[{{B{\"o}hringer} {et~al.}(1994){B{\"o}hringer}, {Briel}, {Schwarz},
  {Voges}, {Hartner}, \& {Tr{\"u}mper}}]{1994Natur.368..828B}
{B{\"o}hringer}, H., {Briel}, U.~G., {Schwarz}, R.~A., {Voges}, W., {Hartner},
  G., \& {Tr{\"u}mper}, J. 1994, \nat, 368, 828

\bibitem[{{Br{\"u}ggen} \& {Kaiser}(2002)}]{2002Natur.418..301B}
{Br{\"u}ggen}, M., \& {Kaiser}, C.~R. 2002, \nat, 418, 301

\bibitem[{{Capelo} {et~al.}(2012){Capelo}, {Coppi}, \&
  {Natarajan}}]{2012MNRAS.422..686C}
{Capelo}, P.~R., {Coppi}, P.~S., \& {Natarajan}, P. 2012, \mnras, 422, 686

\bibitem[{{Cavagnolo} {et~al.}(2009){Cavagnolo}, {Donahue}, {Voit}, \&
  {Sun}}]{2009ApJS..182...12C}
{Cavagnolo}, K.~W., {Donahue}, M., {Voit}, G.~M., \& {Sun}, M. 2009, \apjs,
  182, 12

\bibitem[{{Chandran} \& {Rasera}(2007)}]{2007ApJ...671.1413C}
{Chandran}, B.~D.~G., \& {Rasera}, Y. 2007, \apj, 671, 1413

\bibitem[{{Churazov} {et~al.}(2001){Churazov}, {Br{\"u}ggen}, {Kaiser},
  {B{\"o}hringer}, \& {Forman}}]{2001ApJ...554..261C}
{Churazov}, E., {Br{\"u}ggen}, M., {Kaiser}, C.~R., {B{\"o}hringer}, H., \&
  {Forman}, W. 2001, \apj, 554, 261

\bibitem[{{Churazov} {et~al.}(2008){Churazov}, {Forman}, {Vikhlinin},
  {Tremaine}, {Gerhard}, \& {Jones}}]{2008MNRAS.388.1062C}
{Churazov}, E., {Forman}, W., {Vikhlinin}, A., {Tremaine}, S., {Gerhard}, O.,
  \& {Jones}, C. 2008, \mnras, 388, 1062

\bibitem[{{Churazov} {et~al.}(2010){Churazov}, {Tremaine}, {Forman}, {Gerhard},
  {Das}, {Vikhlinin}, {Jones}, {B{\"o}hringer}, \&
  {Gebhardt}}]{2010MNRAS.404.1165C}
{Churazov}, E., {et~al.} 2010, \mnras, 404, 1165

\bibitem[{{Conroy} \& {Ostriker}(2008)}]{2008ApJ...681..151C}
{Conroy}, C., \& {Ostriker}, J.~P. 2008, \apj, 681, 151

\bibitem[{{Croton} {et~al.}(2006){Croton}, {Springel}, {White}, {De Lucia},
  {Frenk}, {Gao}, {Jenkins}, {Kauffmann}, {Navarro}, \&
  {Yoshida}}]{2006MNRAS.365...11C}
{Croton}, D.~J., {et~al.} 2006, \mnras, 365, 11

\bibitem[{{de Gasperin} {et~al.}(2012){de Gasperin}, {Orr{\'u}}, {Murgia},
  {Merloni}, {Falcke}, {Beck}, {Beswick}, {B{\^i}rzan}, {Bonafede},
  {Br{\"u}ggen}, {Brunetti}, {Chy{\.z}y}, {Conway}, {Croston}, {En{\ss}lin},
  {Ferrari}, {Heald}, {Heidenreich}, {Jackson}, {Macario}, {McKean}, {Miley},
  {Morganti}, {Offringa}, {Pizzo}, {Rafferty}, {R{\"o}ttgering}, {Shulevski},
  {Steinmetz}, {Tasse}, {van der Tol}, {van Driel}, {van Weeren}, {van
  Zwieten}, {Alexov}, {Anderson}, {Asgekar}, {Avruch}, {Bell}, {Bell},
  {Bentum}, {Bernardi}, {Best}, {Breitling}, {Broderick}, {Butcher}, {Ciardi},
  {Dettmar}, {Eisloeffel}, {Frieswijk}, {Gankema}, {Garrett}, {Gerbers},
  {Griessmeier}, {Gunst}, {Hassall}, {Hessels}, {Hoeft}, {Horneffer},
  {Karastergiou}, {K{\"o}hler}, {Koopman}, {Kuniyoshi}, {Kuper}, {Maat},
  {Mann}, {Mevius}, {Mulcahy}, {Munk}, {Nijboer}, {Noordam}, {Paas}, {Pandey},
  {Pandey}, {Polatidis}, {Reich}, {Schoenmakers}, {Sluman}, {Smirnov}, {Sobey},
  {Stappers}, {Swinbank}, {Tagger}, {Tang}, {van Bemmel}, {van Cappellen}, {van
  Duin}, {van Haarlem}, {van Leeuwen}, {Vermeulen}, {Vocks}, {White}, {Wise},
  {Wucknitz}, \& {Zarka}}]{2012A&A...547A..56D}
{de Gasperin}, F., {et~al.} 2012, \aap, 547, A56

\bibitem[{{Dubois} {et~al.}(2011){Dubois}, {Devriendt}, {Teyssier}, \&
  {Slyz}}]{2011MNRAS.417.1853D}
{Dubois}, Y., {Devriendt}, J., {Teyssier}, R., \& {Slyz}, A. 2011, \mnras, 417,
  1853

\bibitem[{{Dursi}(2007)}]{2007ApJ...670..221D}
{Dursi}, L.~J. 2007, \apj, 670, 221

\bibitem[{{Dursi} \& {Pfrommer}(2008)}]{2008ApJ...677..993D}
{Dursi}, L.~J., \& {Pfrommer}, C. 2008, \apj, 677, 993

\bibitem[{{En{\ss}lin} {et~al.}(2011){En{\ss}lin}, {Pfrommer}, {Miniati}, \&
  {Subramanian}}]{2011A&A...527A..99E}
{En{\ss}lin}, T., {Pfrommer}, C., {Miniati}, F., \& {Subramanian}, K. 2011,
  \aap, 527, A99

\bibitem[{{En{\ss}lin} {et~al.}(2007){En{\ss}lin}, {Pfrommer}, {Springel}, \&
  {Jubelgas}}]{2007A&A...473...41E}
{En{\ss}lin}, T.~A., {Pfrommer}, C., {Springel}, V., \& {Jubelgas}, M. 2007,
  \aap, 473, 41

\bibitem[{{Field}(1965)}]{1965ApJ...142..531F}
{Field}, G.~B. 1965, \apj, 142, 531

\bibitem[{{Fujita} \& {Ohira}(2012)}]{2012ApJ...746...53F}
{Fujita}, Y., \& {Ohira}, Y. 2012, \apj, 746, 53

\bibitem[{{Gaspari} {et~al.}(2012{\natexlab{a}}){Gaspari}, {Brighenti}, \&
  {Temi}}]{2012MNRAS.424..190G}
{Gaspari}, M., {Brighenti}, F., \& {Temi}, P. 2012{\natexlab{a}}, \mnras, 424,
  190

\bibitem[{{Gaspari} {et~al.}(2012{\natexlab{b}}){Gaspari}, {Ruszkowski}, \&
  {Sharma}}]{2012ApJ...746...94G}
{Gaspari}, M., {Ruszkowski}, M., \& {Sharma}, P. 2012{\natexlab{b}}, \apj, 746,
  94

\bibitem[{{Gilkis} \& {Soker}(2012)}]{2012MNRAS.427.1482G}
{Gilkis}, A., \& {Soker}, N. 2012, \mnras, 427, 1482

\bibitem[{{Guo} \& {Oh}(2008)}]{2008MNRAS.384..251G}
{Guo}, F., \& {Oh}, S.~P. 2008, \mnras, 384, 251

\bibitem[{{Hofmann}(2006)}]{2006astro.ph..3076H}
{Hofmann}, W. 2006, arXiv:astro-ph/0603076

\bibitem[{{Hudson} {et~al.}(2010){Hudson}, {Mittal}, {Reiprich}, {Nulsen},
  {Andernach}, \& {Sarazin}}]{2010A&A...513A..37H}
{Hudson}, D.~S., {Mittal}, R., {Reiprich}, T.~H., {Nulsen}, P.~E.~J.,
  {Andernach}, H., \& {Sarazin}, C.~L. 2010, \aap, 513, A37

\bibitem[{{Jaffe} \& {Perola}(1973)}]{1973A&A....26..423J}
{Jaffe}, W.~J., \& {Perola}, G.~C. 1973, \aap, 26, 423

\bibitem[{{Kim} \& {Narayan}(2003)}]{2003ApJ...596L.139K}
{Kim}, W.-T., \& {Narayan}, R. 2003, \apjl, 596, L139

\bibitem[{{Komissarov} \& {Gubanov}(1994)}]{1994A&A...285...27K}
{Komissarov}, S.~S., \& {Gubanov}, A.~G. 1994, \aap, 285, 27

\bibitem[{{Kulsrud} \& {Pearce}(1969)}]{1969ApJ...156..445K}
{Kulsrud}, R., \& {Pearce}, W.~P. 1969, \apj, 156, 445

\bibitem[{{Kulsrud}(2005)}]{Kulsrud}
{Kulsrud}, R.~M. 2005, {Plasma physics for astrophysics}

\bibitem[{{Loewenstein} {et~al.}(1991){Loewenstein}, {Zweibel}, \&
  {Begelman}}]{1991ApJ...377..392L}
{Loewenstein}, M., {Zweibel}, E.~G., \& {Begelman}, M.~C. 1991, \apj, 377, 392

\bibitem[{{Lyutikov}(2006)}]{2006MNRAS.373...73L}
{Lyutikov}, M. 2006, \mnras, 373, 73

\bibitem[{{Matsushita} {et~al.}(2002){Matsushita}, {Belsole}, {Finoguenov}, \&
  {B{\"o}hringer}}]{2002A&A...386...77M}
{Matsushita}, K., {Belsole}, E., {Finoguenov}, A., \& {B{\"o}hringer}, H. 2002,
  \aap, 386, 77

\bibitem[{{McCourt} {et~al.}(2012){McCourt}, {Sharma}, {Quataert}, \&
  {Parrish}}]{2012MNRAS.419.3319M}
{McCourt}, M., {Sharma}, P., {Quataert}, E., \& {Parrish}, I.~J. 2012, \mnras,
  419, 3319

\bibitem[{{McNamara} \& {Nulsen}(2007)}]{2007ARA&A..45..117M}
{McNamara}, B.~R., \& {Nulsen}, P.~E.~J. 2007, \araa, 45, 117

\bibitem[{{McNamara} \& {Nulsen}(2012)}]{2012NJPh...14e5023M}
---. 2012, New Journal of Physics, 14, 055023

\bibitem[{{Mei} {et~al.}(2007){Mei}, {Blakeslee}, {C{\^o}t{\'e}}, {Tonry},
  {West}, {Ferrarese}, {Jord{\'a}n}, {Peng}, {Anthony}, \&
  {Merritt}}]{2007ApJ...655..144M}
{Mei}, S., {et~al.} 2007, \apj, 655, 144

\bibitem[{{Million} {et~al.}(2010){Million}, {Werner}, {Simionescu}, {Allen},
  {Nulsen}, {Fabian}, {B{\"o}hringer}, \& {Sanders}}]{2010MNRAS.407.2046M}
{Million}, E.~T., {Werner}, N., {Simionescu}, A., {Allen}, S.~W., {Nulsen},
  P.~E.~J., {Fabian}, A.~C., {B{\"o}hringer}, H., \& {Sanders}, J.~S. 2010,
  \mnras, 407, 2046

\bibitem[{{Narayan} \& {Medvedev}(2001)}]{2001ApJ...562L.129N}
{Narayan}, R., \& {Medvedev}, M.~V. 2001, \apjl, 562, L129

\bibitem[{{Nulsen} {et~al.}(2005){Nulsen}, {McNamara}, {Wise}, \&
  {David}}]{2005ApJ...628..629N}
{Nulsen}, P.~E.~J., {McNamara}, B.~R., {Wise}, M.~W., \& {David}, L.~P. 2005,
  \apj, 628, 629

\bibitem[{{Owen} {et~al.}(2000){Owen}, {Eilek}, \&
  {Kassim}}]{2000ApJ...543..611O}
{Owen}, F.~N., {Eilek}, J.~A., \& {Kassim}, N.~E. 2000, \apj, 543, 611

\bibitem[{{Pacholczyk}(1970)}]{1970ranp.book.....P}
{Pacholczyk}, A.~G. 1970, {Radio astrophysics. Nonthermal processes in galactic
  and extragalactic sources}

\bibitem[{{Parrish} \& {Quataert}(2008)}]{2008ApJ...677L...9P}
{Parrish}, I.~J., \& {Quataert}, E. 2008, \apjl, 677, L9

\bibitem[{{Parrish} {et~al.}(2010){Parrish}, {Quataert}, \&
  {Sharma}}]{2010ApJ...712L.194P}
{Parrish}, I.~J., {Quataert}, E., \& {Sharma}, P. 2010, \apjl, 712, L194

\bibitem[{{Peterson} \& {Fabian}(2006)}]{2006PhR...427....1P}
{Peterson}, J.~R., \& {Fabian}, A.~C. 2006, \physrep, 427, 1

\bibitem[{{Peterson} {et~al.}(2003){Peterson}, {Kahn}, {Paerels}, {Kaastra},
  {Tamura}, {Bleeker}, {Ferrigno}, \& {Jernigan}}]{2003ApJ...590..207P}
{Peterson}, J.~R., {Kahn}, S.~M., {Paerels}, F.~B.~S., {Kaastra}, J.~S.,
  {Tamura}, T., {Bleeker}, J.~A.~M., {Ferrigno}, C., \& {Jernigan}, J.~G. 2003,
  \apj, 590, 207

\bibitem[{{Pfrommer} {et~al.}(2012){Pfrommer}, {Chang}, \&
  {Broderick}}]{2012ApJ...752...24P}
{Pfrommer}, C., {Chang}, P., \& {Broderick}, A.~E. 2012, \apj, 752, 24

\bibitem[{{Pfrommer} \& {Dursi}(2010)}]{2010NatPh...6..520P}
{Pfrommer}, C., \& {Dursi}, L.~J. 2010, Nature Physics, 6, 520

\bibitem[{{Pfrommer} \& {En{\ss}lin}(2003)}]{2003A&A...407L..73P}
{Pfrommer}, C., \& {En{\ss}lin}, T.~A. 2003, \aap, 407, L73

\bibitem[{{Pfrommer} \& {En{\ss}lin}(2004)}]{2004A&A...413...17P}
---. 2004, \aap, 413, 17

\bibitem[{{Pfrommer} {et~al.}(2005){Pfrommer}, {En{\ss}lin}, \&
  {Sarazin}}]{2005A&A...430..799P}
{Pfrommer}, C., {En{\ss}lin}, T.~A., \& {Sarazin}, C.~L. 2005, \aap, 430, 799

\bibitem[{{Pfrommer} {et~al.}(2008){Pfrommer}, {En{\ss}lin}, \&
  {Springel}}]{2008MNRAS.385.1211P}
{Pfrommer}, C., {En{\ss}lin}, T.~A., \& {Springel}, V. 2008, \mnras, 385, 1211

\bibitem[{{Pinzke} {et~al.}(2011){Pinzke}, {Pfrommer}, \&
  {Bergstr{\"o}m}}]{2011PhRvD..84l3509P}
{Pinzke}, A., {Pfrommer}, C., \& {Bergstr{\"o}m}, L. 2011, \prd, 84, 123509

\bibitem[{{Pizzolato} \& {Soker}(2005)}]{2005ApJ...632..821P}
{Pizzolato}, F., \& {Soker}, N. 2005, \apj, 632, 821

\bibitem[{{Puchwein} \& {Springel}(2013)}]{2013MNRAS.428.2966P}
{Puchwein}, E., \& {Springel}, V. 2013, \mnras, 428, 2966

\bibitem[{{Quataert}(2008)}]{2008ApJ...673..758Q}
{Quataert}, E. 2008, \apj, 673, 758

\bibitem[{{Reimer} {et~al.}(2003){Reimer}, {Pohl}, {Sreekumar}, \&
  {Mattox}}]{2003ApJ...588..155R}
{Reimer}, O., {Pohl}, M., {Sreekumar}, P., \& {Mattox}, J.~R. 2003, \apj, 588,
  155

\bibitem[{{Rieger} \& {Aharonian}(2012)}]{2012MPLA...2730030R}
{Rieger}, F.~M., \& {Aharonian}, F. 2012, Modern Physics Letters A, 27, 30030

\bibitem[{{Romanowsky} \& {Kochanek}(2001)}]{2001ApJ...553..722R}
{Romanowsky}, A.~J., \& {Kochanek}, C.~S. 2001, \apj, 553, 722

\bibitem[{{Ruszkowski} \& {Begelman}(2002)}]{2002ApJ...581..223R}
{Ruszkowski}, M., \& {Begelman}, M.~C. 2002, \apj, 581, 223

\bibitem[{{Ruszkowski} {et~al.}(2004){Ruszkowski}, {Br{\"u}ggen}, \&
  {Begelman}}]{2004ApJ...611..158R}
{Ruszkowski}, M., {Br{\"u}ggen}, M., \& {Begelman}, M.~C. 2004, \apj, 611, 158

\bibitem[{{Ruszkowski} {et~al.}(2007){Ruszkowski}, {En{\ss}lin}, {Br{\"u}ggen},
  {Heinz}, \& {Pfrommer}}]{2007MNRAS.378..662R}
{Ruszkowski}, M., {En{\ss}lin}, T.~A., {Br{\"u}ggen}, M., {Heinz}, S., \&
  {Pfrommer}, C. 2007, \mnras, 378, 662

\bibitem[{{Ruszkowski} \& {Oh}(2010)}]{2010ApJ...713.1332R}
{Ruszkowski}, M., \& {Oh}, S.~P. 2010, \apj, 713, 1332

\bibitem[{{Sanders} \& {Fabian}(2007)}]{2007MNRAS.381.1381S}
{Sanders}, J.~S., \& {Fabian}, A.~C. 2007, \mnras, 381, 1381

\bibitem[{{Sanders} {et~al.}(2008){Sanders}, {Fabian}, {Allen}, {Morris},
  {Graham}, \& {Johnstone}}]{2008MNRAS.385.1186S}
{Sanders}, J.~S., {Fabian}, A.~C., {Allen}, S.~W., {Morris}, R.~G., {Graham},
  J., \& {Johnstone}, R.~M. 2008, \mnras, 385, 1186

\bibitem[{{Sanders} {et~al.}(2010){Sanders}, {Fabian}, {Frank}, {Peterson}, \&
  {Russell}}]{2010MNRAS.402..127S}
{Sanders}, J.~S., {Fabian}, A.~C., {Frank}, K.~A., {Peterson}, J.~R., \&
  {Russell}, H.~R. 2010, \mnras, 402, 127

\bibitem[{{Sanders} {et~al.}(2009){Sanders}, {Fabian}, \&
  {Taylor}}]{2009MNRAS.393...71S}
{Sanders}, J.~S., {Fabian}, A.~C., \& {Taylor}, G.~B. 2009, \mnras, 393, 71

\bibitem[{{Sharma} {et~al.}(2009){Sharma}, {Chandran}, {Quataert}, \&
  {Parrish}}]{2009ApJ...699..348S}
{Sharma}, P., {Chandran}, B.~D.~G., {Quataert}, E., \& {Parrish}, I.~J. 2009,
  \apj, 699, 348

\bibitem[{{Sharma} {et~al.}(2012){Sharma}, {McCourt}, {Quataert}, \&
  {Parrish}}]{2012MNRAS.420.3174S}
{Sharma}, P., {McCourt}, M., {Quataert}, E., \& {Parrish}, I.~J. 2012, \mnras,
  420, 3174

\bibitem[{{Sijacki} {et~al.}(2008){Sijacki}, {Pfrommer}, {Springel}, \&
  {En{\ss}lin}}]{2008MNRAS.387.1403S}
{Sijacki}, D., {Pfrommer}, C., {Springel}, V., \& {En{\ss}lin}, T.~A. 2008,
  \mnras, 387, 1403

\bibitem[{{Sutherland} \& {Dopita}(1993)}]{1993ApJS...88..253S}
{Sutherland}, R.~S., \& {Dopita}, M.~A. 1993, \apjs, 88, 253

\bibitem[{{Vazza} {et~al.}(2013){Vazza}, {Br{\"u}ggen}, \&
  {Gheller}}]{2013MNRAS.428.2366V}
{Vazza}, F., {Br{\"u}ggen}, M., \& {Gheller}, C. 2013, \mnras, 428, 2366

\bibitem[{{Voigt} \& {Fabian}(2004)}]{2004MNRAS.347.1130V}
{Voigt}, L.~M., \& {Fabian}, A.~C. 2004, \mnras, 347, 1130

\bibitem[{{Voit}(2011)}]{2011ApJ...740...28V}
{Voit}, G.~M. 2011, \apj, 740, 28

\bibitem[{{Wagner} {et~al.}(2009){Wagner}, {Lindfors}, {Sillanp{\"a}{\"a}},
  {Wagner}, \& {CTA Consortium}}]{2009arXiv0912.3742W}
{Wagner}, R.~M., {Lindfors}, E.~J., {Sillanp{\"a}{\"a}}, A., {Wagner}, S., \&
  {CTA Consortium}. 2009, arXiv:0912.3742

\bibitem[{{Wentzel}(1971)}]{1971ApJ...163..503W}
{Wentzel}, D.~G. 1971, \apj, 163, 503

\bibitem[{{Werner} {et~al.}(2010){Werner}, {Simionescu}, {Million}, {Allen},
  {Nulsen}, {von der Linden}, {Hansen}, {B{\"o}hringer}, {Churazov}, {Fabian},
  {Forman}, {Jones}, {Sanders}, \& {Taylor}}]{2010MNRAS.407.2063W}
{Werner}, N., {et~al.} 2010, \mnras, 407, 2063

\bibitem[{{Werner} {et~al.}(2013){Werner}, {Oonk}, {Canning}, {Allen},
  {Simionescu}, {Kos}, {van Weeren}, {Edge}, {Fabian}, {von der Linden},
  {Nulsen}, {Reynolds}, \& {Ruszkowski}}]{2013ApJ...767..153W}
---. 2013, \apj, 767, 153

\bibitem[{{Wiener} {et~al.}(2013{\natexlab{a}}){Wiener}, {Oh}, \&
  {Guo}}]{2013MNRAS.434.2209W}
{Wiener}, J., {Oh}, S.~P., \& {Guo}, F. 2013{\natexlab{a}}, \mnras, 434, 2209

\bibitem[{{Wiener} {et~al.}(2013{\natexlab{b}}){Wiener}, {Zweibel}, \&
  {Oh}}]{2013ApJ...767...87W}
{Wiener}, J., {Zweibel}, E.~G., \& {Oh}, S.~P. 2013{\natexlab{b}}, \apj, 767,
  87

\bibitem[{{Zakamska} \& {Narayan}(2003)}]{2003ApJ...582..162Z}
{Zakamska}, N.~L., \& {Narayan}, R. 2003, \apj, 582, 162

\end{thebibliography}
\bibliographystyle{apj}

\end{document}